\documentclass[pra,twocolumn,showpacs,preprintnumbers,amsmath,amssymb]{revtex4}

\usepackage{graphicx}
\usepackage{dcolumn}
\usepackage{bm}

\DeclareMathOperator{\sinc}{sinc}
\DeclareMathOperator{\cotan}{cotan}
\DeclareMathOperator{\cotanh}{cotanh}

\begin{document}

\title{Photon noise in a random laser amplifier with fluctuating properties}

\author{V.Yu. Fedorov}
\affiliation{Laboratoire de Physique et Mod\'{e}lisation des Milieux Condens\'{e}s, Universit\'{e} Joseph Fourier and CNRS, 25 rue des Martyrs, 38042 Grenoble, France}

\author{S.E. Skipetrov}
\affiliation{Laboratoire de Physique et Mod\'{e}lisation des Milieux Condens\'{e}s, Universit\'{e} Joseph Fourier and CNRS, 25 rue des Martyrs, 38042 Grenoble, France}

\date{\today}

\begin{abstract}
We study fluctuations of the number of photocounts measured by an ideal photodetector illuminated by light scattered in an amplifying disordered medium, below the threshold for random lasing. We show that the variance of fluctuations and their correlation function carry information about fluctuating properties of the medium. A direct link is established between the fluctuations of the number of photocounts due to the amplified spontaneous emission (ASE) and the dimensionless conductance $g$ of the medium. Our results suggest a possibility of probing amplifying disordered media by analyzing statistics of their ASE, without illuminating them from outside by a probe beam. 
\end{abstract}


\maketitle

\section{Introduction}
The interplay of multiple scattering and amplification can lead to interesting and still poorly understood phenomena, culminating in ``random lasing'' \cite{cao05,wiersma08}. The random laser is a strongly scattering disordered medium where losses due to leakage of light through open boundaries are compensated by amplification inside the medium (sustained due to some external pump mechanism), so that lasing condition is fulfilled despite the absence of an external cavity \cite{letokhov68}. Similar to conventional lasers (i.e. lasers composed of a transparent amplifying medium inside a high-Q cavity) \cite{haken85,siegman86}, random lasers start laser emission only when the pump power reaches a threshold. Below the threshold, the system represents a ``random laser amplifier'': it amplifies light propagating inside it (though the amplification is not strong enough to compensate for losses) \footnote{In a random laser amplifier, ``amplification'' means that the total (i.e. integrated over all directions, both in transmission and in reflection) power leaving the system exceeds the incident power.} and it is expected to exhibit properties similar to that of conventional laser amplifiers \cite{mandel95}: increased noise in transmission, amplified spontaneous emission, etc. Random laser amplifiers are promising systems for studies of Anderson localization of light \cite{anderson58,abrahams79,john91} and, more generally, any interference effects in disordered media because, on the one hand, the amplification gives increased weight to long-lived quasi-modes (or, equivalently, to long wave paths inside the medium) and, on the other hand, the system remains linear, in contrast to ``true'' random lasers, operating above threshold in strongly nonlinear regimes.    
However, precisely because they operate below laser threshold, laser amplifiers require full quantum treatment that properly accounts for spontaneous emission processes \cite{haken85,siegman86,mandel95}. One of the quantities for which such a treatment is crucial is the photocount probability distribution $p(n)$, giving the probability density of counting $n$ photons at the output of the amplifier during a given sampling time $\tau$.       
In the classical picture, $p(n)$ would be akin to that in a medium without amplification, with a larger average value of $n$, but with exactly the same values of normalized higher statistical moments. In the quantum picture, which corresponds to the physical reality observed in experiments, laser amplifiers are known to add noise to the incident light and to generate a noisy light field (amplified spontaneous emission, ASE) even in the absence of an external signal \cite{haken85,siegman86,mandel95}. For random laser amplifiers with time-independent properties (quenched disorder) $p(n)$ was studied by Beenakker and Patra in a series of papers \cite{bee98,patra99b,patra99} using the methods of random-matrix theory. More recently, Florescu and John have considered photon statistics of light emitted by random lasers \cite{florescu04}. 

In the present paper we study the first two statistical moments of $p(n)$ for light transmitted through or emitted by random laser amplifiers with \emph{fluctuating} properties. A simple example of such a system is a suspension of scattering centers (particles) in Brownian motion, with amplification either in the background medium (typically, a laser dye) or in the particles themselves. Such media are often used in experiments on random lasers \cite{lawandy94,wiersma95,noginov95,cao00,soest01,dice05,wu06}. Because of the slow motion of scattering centers, all possible configurations of disorder are explored continuosly, leading to continuous changes of transmission and reflection coefficients of the medium. As a consequence, two simple limits exist. First, for short sampling times $\tau$, well below the correlation time $t_c$ of, say, transmission coefficients (if the measurement is performed in transmission), the measurement yields the same result as if the scattering centers were immobile (quenched disorder). In the opposite limit of very long sampling times $\tau$, much in excess of $t_c$, configurational average is effectively performed due to the scatterer motion and the measurement directly yields the ensemble average of the measured quantity (annealed disorder). However, only in these two limits a direct link between the results obtained for stationary and fluctuating media can be established. In the general case of arbitrary $\tau$ and, in particular, when $\tau \sim t_c$, the result will depend on the particular type of scatterer dynamics and cannot be predicted from \textit{a priori} considerations. It is precisely this general situation that we consider in the present paper. Our analysis is rather general and applies to any random laser amplifier with fluctuating properties. To be more specific, we illustrate our main results on the particularly important example of a liquid suspension of scattering particles in Brownian motion, where amplification can take place either in the particles, in the liquid, or in both.

\section{Input-output relations}

\begin{figure}
\begin{flushright}
\hspace*{9mm}\includegraphics[width=\columnwidth]{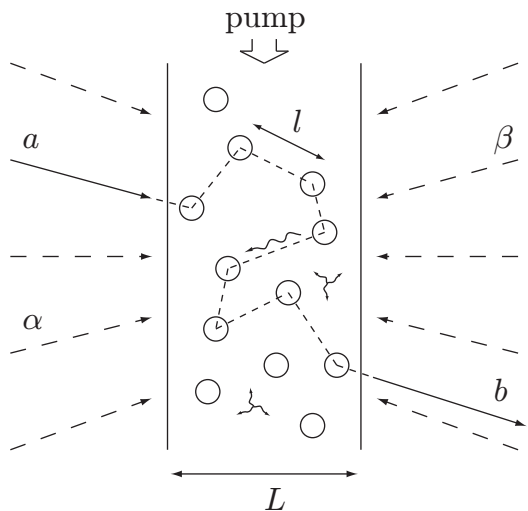}
\end{flushright}
\caption{\label{fig:Media}
A slab-shaped sample of thickness $L$ is filled with an amplifying disordered medium represented by discrete, mobile scattering centers (circles) with amplification of light either in the scatterers, in the surrounding medium, or in both. A spatially uniform amplification is achieved through an external pump mechanism. The slab is illuminated by a probe light in the incoming mode $a$; all other incoming modes $\alpha$ and $\beta$ (dashed arrows) are in vacuum states. We are interested in the fluctuations of transmitted light in the mode $b$. The polygonal dashed path inside the slab illustrates multiple scattering of light with a mean free path $\ell \ll L$. Wavy lines with arrows symbolize stimulated and spontaneous emission of light by the medium.}
\end{figure}

We consider a slab of amplifying disordered medium with time-dependent properties (see Fig.~\ref{fig:Media}). The slab has thickness $L$, much exceeding the mean free path $\ell$ due to scattering, and cross-sectional area $A$. The mean free path is assumed to far exceed the wavelength $\lambda$ of light under consideration. At each frequency $\omega$, incident and scattered waves can be decomposed over sets of $2 N(\omega) = 4 \times k^2 A/4 \pi \gg 1$ transverse modes supported by the slab. Here $k = \omega/c$ and $c$ is the speed of light in the free space. In counting the number of transverse modes we have taken into account 2 propagation directions (from left to right and from right to left) and 2 polarization states of the electromagnetic field. In our notation $N(\omega)$ will stay for the number of left (or right) propagating modes at each side of the slab.  We will use subscripts $\alpha$ and $\beta$ for incoming modes, incident on the slab from the left and from the right, respectively. They will take values $\alpha = 1, \ldots, N(\omega)$ and $\beta = N(\omega) + 1, \ldots, 2N(\omega)$. The outgoing modes will be numbered by $2N(\omega) + 1, ..., 3N(\omega)$ for right-propagating modes and by $3N(\omega) + 1, ..., 4N(\omega)$ for left-propagating ones. The light incident on the slab is assumed to be quasi-monochromatic and corresponds to a single mode $\alpha = a$. It can be in arbitrary quantum state. All other incoming modes are assumed to be in vacuum states. Our purpose will be the analysis of photon statistics in the outgoing mode $b$ on the right of the random medium (see Fig.\ \ref{fig:Media}) in the presence of a pump mechanism that ensures amplification of light inside the slab. The amplification is assumed to be uniform throughout the slab. It is characterized by the amplification time $t_a$ during which the amplitude of a propagating wave is multiplied by $e$. The corresponding amplification length is  $\ell_a = c t_a$. In the context of this paper, we will not need to specify the physical mechanism through which pumping is achieved.

In order to obtain the quantum-mechanical description of the problem, we need to quantize the electromagnetic field. This can be done in the usual way \cite{mandel95} by replacing the amplitudes of the incoming and outgoing modes $a(\omega)$ and their complex conjugates $a^{*}(\omega)$ by annihilation and creation operators ${\hat a}(\omega)$ and ${\hat a}^{\dagger}(\omega)$, respectively. These operators obey the usual bosonic commutation relations,
\begin{equation}
[\hat{a}_{i}\left(\omega\right),\hat{a}_{j}^{\dagger}\left(\omega'\right)] =\delta_{ij}\delta(\omega-\omega')
\label{eq:CommutationsFreaq}
\end{equation}
where $i$ and $j$ run over all incoming or all outgoing modes.
Input-output relations between operators corresponding to incoming and outgoing modes in the case of an amplifying disordered slab with \emph{time-independent} properties have been established in Refs.\ \cite{bee98, patra99b, patra99, vivi03}:
\begin{eqnarray}
\hat{a}_{b}\left(\omega\right) &=&
\sum_{\alpha = 1}^{N(\omega)} t_{\alpha b}\left(\omega\right)\hat{a}_{\alpha}\left(\omega\right)+
\sum_{\beta = N(\omega)+1}^{2N(\omega)} r_{\beta b}\left(\omega\right)\hat{a}_{\beta}\left(\omega\right)
\nonumber \\
&+& \sum_{\gamma}v_{\gamma b}^{*}\left(\omega\right)\hat{c}_{\gamma}^{\dagger}
\left(\omega\right)
\label{ioimmobile}
\end{eqnarray}
Here $t_{\alpha b}$ and $r_{\beta b}$ are transmission and reflection coefficients from the incoming modes $\alpha$ and $\beta$, respectively, to the outgoing mode $b$. The last sum in Eq.\ (\ref{ioimmobile}) describes spontaneous emission of light in the amplifying medium, with $\gamma$ running over ``objects'' (e.g., atoms or molecules) emitting at frequency $\omega$. To be precise, the phenomenological model leading to Eq.\ (\ref{ioimmobile}) is based on the assumption that the interaction of light with a linear amplifying medium can be modeled by coupling of optical modes to a bath of inverted harmonic oscillators \cite{gardiner00}. The number of terms in the last sum of Eq.\ (\ref{ioimmobile}) is then the number of oscillators in the bath, whereas the bosonic operators ${\hat c}_{\gamma}$ obey
\begin{eqnarray}
&&[\hat{c}_{i}(\omega),\hat{c}_{j}^{\dagger}(\omega')] = \delta_{ij}\delta(\omega-\omega')
\label{commc}
\\
&&\langle \hat{c}_{i}(\omega) \hat{c}_{j}^{\dagger}(\omega') \rangle = -\delta_{ij}\delta(\omega-\omega') \eta(\omega)
\label{corrc}
\end{eqnarray}
with $\eta(\omega) = [\exp(\hbar \omega/k T) - 1]^{-1}$ and negative effective temperature $T < 0$.

In a medium with \emph{time-dependent} properties the input-output relations (\ref{ioimmobile}) should be generalized and become
\begin{eqnarray}
\hat{a}_{b}\left(\omega\right) &=& \int_{-\infty}^{\infty} d \omega'
\left[ \sum\limits_{\alpha=1}^{N(\omega')}
{\tilde t}_{\alpha b}\left(\omega,\omega'\right)\hat{a}_{\alpha}\left(\omega'\right)
\right. \nonumber \\
&+& \left. \sum\limits_{\beta=N(\omega')+1}^{2N(\omega')}{\tilde r}_{\beta b}\left(\omega, \omega'\right)\hat{a}_{\beta}\left(\omega'\right)
\right. \nonumber \\
&+& \left. \sum\limits_{\gamma} {\tilde v}_{\gamma b}^{*}\left(\omega,\omega'\right)\hat{c}_{\gamma}^{\dagger}\left(\omega'\right)
\right]
\label{eq:GeneralIOrelation}
\end{eqnarray}
Modes at different frequencies are not independent anymore: the incident field at a given frequency $\omega^{\prime}$ can give rise to outgoing waves at different frequencies $\omega \ne \omega^{\prime}$. The physical meaning of this can be understood on the simple example of a suspension of scattering particles. Because of the thermal motion of particles, scattered light will suffer from Doppler broadening and hence will not be monochromatic even if the incident light were. The Doppler broadening of transmitted, reflected, and spontaneously emitted light are described by the coefficients ${\tilde t}_{\alpha b}(\omega,\omega')$, ${\tilde r}_{\beta b}(\omega,\omega')$ and ${\tilde v}_{\gamma b}^{*}(\omega,\omega')$, respectively.

\begin{figure}
\begin{centering}
\includegraphics[width=\columnwidth]{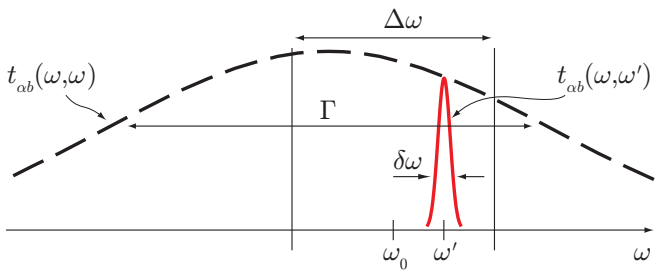}
\end{centering}
\caption{\label{fig:FreaquencyScales}
Frequency scales of the problem. 
The wide dashed curve schematizes the slow dependence of the transmission coefficient $t_{\alpha b}(\omega,\omega')$ on $\omega = \omega'$. The narrow solid curve depicts the dependence of this coefficient on $\omega \ne \omega'$. When $\omega'$ is varied, the latter curve shifts along $\omega$ axis, keeping its overall shape and remaining centered at $\omega'$. Its maximum touches the curve $t_{\alpha b}(\omega,\omega)$. The width $\delta \omega$ of the solid curve is much smaller than the width $\Gamma$ of the dashed curve. A bandpass spectral filter limits any measurement to a frequency band centered at $\omega_0$. The pass band of the filter $\Delta \omega$ obeys $\delta \omega \ll \Delta\omega \ll \Gamma$.}
\end{figure}

Let us now show that Eq.\ (\ref{eq:GeneralIOrelation}) indeed describes a medium with time-dependent properties. To this end we consider a realistic situation of a measurement device sensitive only to light within a narrow frequency band $\Delta \omega \ll \omega_0$ around the frequency $\omega_0$ of the incident wave. The bandwidth $\Delta \omega$ can be due either to some technical limitations of the device or to a band-pass filter put in front of it on purpose.
Taking spectral filtering into account, the result of any measurement performed on the outgoing mode $b$ can be expressed through an inverse windowed Fourier transform ${\hat a}_b(t)$ of ${\hat a}_b(\omega)$:
\begin{eqnarray}
{\hat a}_b(t) = \frac{1}{\sqrt{2\pi}} \int_{\omega_0 - \Delta\omega/2}^{\omega_0 + \Delta\omega/2} d\omega e^{-i \omega t}
{\hat a}_b(\omega) 
\label{wft}
\end{eqnarray} 
An expression for ${\hat a}_b(t)$ can be obtained from Eq.\ (\ref{eq:GeneralIOrelation}) by applying the inverse windowed Fourier transform to its right-hand side. For the first sum of Eq.\ (\ref{eq:GeneralIOrelation}), for example, this gives 
\begin{eqnarray}
&&\frac{1}{\sqrt{2\pi}}\int_{\omega_0 - \Delta\omega/2}^{\omega_0 + \Delta\omega/2} d\omega e^{-i \omega t}
\nonumber \\
&&\hspace{10mm} \int_{-\infty}^{\infty} d\omega'
\sum\limits_{\alpha=1}^{N(\omega')}
{\tilde t}_{\alpha b}\left(\omega,\omega'\right)\hat{a}_{\alpha}\left(\omega'\right)
\label{wft1}
\end{eqnarray} 
We now introduce sum and difference variables ${\bar \omega} = (\omega + \omega')/2$ and $\Omega = \omega-\omega'$ and define a new transmission coefficient depending on these variables:
$t_{\alpha b}({\bar \omega}, \Omega) = {\tilde t}_{\alpha b}(\omega, \omega')$. As a function of $\Omega$, the latter will be assumed to be different from zero only in a narrow frequency band $\delta \omega \ll \Delta \omega$ around $\Omega = 0$, which corresponds to slow variation of the properties of the random medium on the time scale of $1/\Delta \omega$. As a consequence, not only the integration over $\omega$, but also the integration over $\omega'$ will be effectively limited to the pass band $\Delta \omega$ of the filter. We can therefore replace $N(\omega')$ in the upper limit of summation in Eq.\ (\ref{wft1}) by $N(\omega_0)$. To lighten the notation we will drop the argument $\omega_0$ of $N(\omega_0)$ in the following and will write $N = N(\omega_0)$. Next, we assume that as a function of ${\bar \omega}$, $t_{\alpha b}$ vary significantly only on the scale of $\Gamma \gg \Delta \omega$. Then, because $\omega$, $\omega'$ and hence ${\bar \omega}$ remain within the pass band of the filter, we can replace the first argument ${\bar \omega}$ of $t_{\alpha b}$ by $\omega_0$ and obtain
\begin{eqnarray}
&&\frac{1}{\sqrt{2\pi}}
\sum\limits_{\alpha=1}^{N} \int_{\omega_0 - \Delta\omega/2}^{\omega_0 + \Delta\omega/2} d\omega e^{-i \omega t}
\nonumber \\
&&\hspace{10mm}
\int_{-\infty}^{\infty} d\omega' t_{\alpha b}\left(\omega_0,\omega-\omega'\right)\hat{a}_{\alpha}\left(\omega'\right)
\label{wft2}
\end{eqnarray} 
The next step is to represent $t_{\alpha b}\left(\omega_0,\omega-\omega'\right)$ as a Fourier transform of $t_{\alpha b}\left(\omega_0, t' \right)$ and to perform the integration over $\omega$. This yields
\begin{eqnarray}
&&\sum\limits_{\alpha=1}^{N} 
\int_{-\infty}^{\infty} d \omega'
\int_{-\infty}^{\infty} d t'
e^{-i \omega' t' - i \omega_0 (t-t')}
\nonumber \\
&&\hspace{10mm}
t_{\alpha b}\left(\omega_0, t'\right)\hat{a}_{\alpha}\left(\omega'\right)
\delta_{\Delta \omega}(t-t')
\label{wft3}
\end{eqnarray} 
where we defined
\begin{eqnarray}
\delta_{\Delta \omega}(u) &=& 
\frac{1}{2\pi} \int_{-\Delta \omega/2}^{\Delta \omega/2} d \omega e^{-i \omega u}
\nonumber \\
&=&
\frac{\Delta \omega}{2\pi} \sinc\left(\frac{\Delta\omega u}{2}\right)
\label{delta}
\end{eqnarray} 
and $\sinc(x) = \sin(x)/x$. $\delta_{\Delta \omega}(u)$ tends to the Dirac delta-function in the limit of $\Delta \omega \rightarrow \infty$.
Because $t_{\alpha b}\left(\omega_0, t'\right)$ is a slowly varying function of its second argument (time) and only vary significantly on the scale of $1/\delta \omega \gg 1/\Delta \omega$, while $\delta_{\Delta \omega}(u)$ is peaked around $t = t'$ and has a width $1/\Delta  \omega$, we can replace $t'$ by $t$ in the argument of $t_{\alpha b}\left(\omega_0, t'\right)$ and take the latter out of the integral in Eq.\ (\ref{wft3}). The remaining integrals over $t'$ and $\omega'$ are easily calculated, leaving us with
\begin{eqnarray}
\sqrt{2 \pi} \sum\limits_{\alpha=1}^{N}
t_{\alpha b}\left(\omega_0, t \right) \hat{a}_{\alpha}\left(t\right)
\label{wft4}
\end{eqnarray}
where $\hat{a}_{\alpha}\left(t\right)$ is the windowed inverse Fourier transform of $\hat{a}_{\alpha}\left(\omega\right)$ defined as in Eq.\ (\ref{wft}). Finally, because the three remaining sums in Eq.\ (\ref{eq:GeneralIOrelation}) can be treated in the same way, we obtain the input-output relations in the time-domain:
\begin{eqnarray}
\hat{a}_{b}\left(t\right) &=& \sum_{\alpha=1}^{N}t_{\alpha b}\left(t\right)\hat{a}_{\alpha}\left(t\right)+
\sum_{\beta=N+1}^{2N}r_{\beta b}\left(t\right)\hat{a}_{\beta}\left(t\right)
\nonumber \\
&+& \sum_{\gamma}v_{\gamma b}^{*}\left(t\right)\hat{c}_{\gamma}^{\dagger}\left(t\right)
\label{eq:GeneralIOrelationTime}
\end{eqnarray}
where $t_{\alpha b}\left(t\right) = \sqrt{2\pi} t_{\alpha b}\left(\omega_0, t\right)$ and similarly for $r_{\beta b}$ and $v_{\gamma b}^*$, except that the Fourier transforms are defined with opposite signs of $i$ for $v_{\gamma b}$ and ${\hat c}_{\gamma}$:
\begin{eqnarray}
v_{\gamma b}^{*}(\omega_0, t) &=& \frac{1}{\sqrt{2\pi}} \int_{-\infty}^{\infty} d\Omega e^{-i \Omega t}
v_{\gamma b}^{*}(\omega_0, \Omega) 
\label{vft}
\\
{\hat c}_{\gamma}^{\dagger}(t) &=& \frac{1}{\sqrt{2\pi}} \int_{\omega_0 - \Delta\omega/2}^{\omega_0 + \Delta\omega/2} d\omega e^{-i \omega t}
{\hat c}_{\gamma}^{\dagger}(\omega) 
\label{cft}
\end{eqnarray} 
We will use the input-output relations (\ref{eq:GeneralIOrelationTime}) in the remainder of the paper. They are valid under the assumption of
$\delta \omega \ll \Delta \omega \ll \Gamma$, where $\delta\omega$  
is the largest of the widths of Fourier transforms of $t_{\alpha b}\left(t\right)$, $r_{\beta b}\left(t\right)$ and $v_{\gamma b}^*\left(t\right)$;
$\Delta \omega$ is the width of the frequency band to which the measurement is limited, and $\Gamma$ is the typical scale of variations of $t_{\alpha b}\left(\omega_0, t\right)$, $r_{\beta b}\left(\omega_0, t\right)$ and $v_{\gamma b}^*\left(\omega_0, t\right)$ as functions of the carrier frequency $\omega_0$. Physically, the first condition $\delta\omega \ll \Delta\omega$ means that we are considering the limit of slow variations of the properties of disordered medium (small $\delta \omega$ and hence long typical times $\sim 1/\delta\omega$), such that the broadening of the spectrum due to these variations is well below $\Delta \omega$. The second condition $\Delta\omega \ll \Gamma$ allows us to neglect variations of the frequency-dependent properties of the medium (such as the index of refraction, the mean free path, etc.) within the pass band $\Delta \omega$. This significantly simplifies calculations and makes the results more general and not limited to a particular random medium. The different time scales of the problem and their hierarchy are illustrated in Fig.\ \ref{fig:FreaquencyScales}.

Working in the time domain requires commutation relation for time-dependent operators. These are easily obtained from Eqs.\ (\ref{eq:CommutationsFreaq}) and (\ref{commc}): 
\begin{eqnarray}
&&[\hat{a}_{i}\left(t\right),\hat{a}_{j}^{\dagger}\left(t'\right)] =
\delta_{ij}\delta_{\Delta\omega}\left(t-t'\right)
e^{-i \omega_0 (t-t')}
\label{commat}
\\
&&[ \hat{c}_{i}\left(t\right),\hat{c}_{j}^{\dagger}\left(t'\right) ] = \delta_{ij}\delta_{\Delta \omega}\left(t-t'\right)
e^{i \omega_0 (t-t')}
\label{eq:CommutationsTime}
\end{eqnarray}
Equation (\ref{corrc}) in time domain becomes
\begin{equation}
\langle \hat{c}_{i}\left(t_{i}\right)\hat{c}_{j}^{\dagger}\left(t_{j}\right)\rangle =-\delta_{ij} \eta(\omega_0)
\delta_{\Delta \omega}(t-t') e^{i \omega_0 (t-t')}
\label{eq:ExpectationTime}
\end{equation}
where we have taken into account the condition $\Delta \omega \ll \omega_0$ and approximated $\eta(\omega)$ by $\eta(\omega_0)$ for all $\omega \in [\omega_0 - \Delta\omega/2, \omega_0 + \Delta\omega/2]$.
Note that $\eta(\omega_0) < 0$ in an amplifying medium.

It will appear convenient in what follows to define \emph{intensity} transmission, reflection, and spontaneous emission coefficients:
$T_{\alpha b}(t) = |t_{\alpha b}(t)|^2$,
$R_{\beta b}(t) = |r_{\beta b}(t)|^2$
and 
$V_{\gamma b}(t) = |v_{\gamma b}(t)|^2$.
Requiring that the commutation relations (\ref{commat}) hold for the outgoing mode $b$ and making use of the slowness of variations of $t_{\alpha b}(t)$, $r_{\beta b}(t)$ and $v_{\gamma b}(t)$ on the scale of $1/\Delta \omega$, we obtain an energy conservation law for our problem:
\begin{equation}
T_{b}\left(t\right)+R_{b}\left(t\right)-V_{b}\left(t\right)=1
\label{eq:ConservationLaw}
\end{equation}
where we defined the \emph{total\/} transmission, reflection, and spontaneous emission coefficients:
$T_{b}(t) = \sum_{\alpha} T_{\alpha b}(t)$,
$R_{b}(t) = \sum_{\beta} R_{\beta b}(t)$ and
$V_{b}(t) = \sum_{\gamma} V_{\gamma b}(t)$.
We note here that these definitions of the total transmission and reflection coefficients $T_b$ and $R_b$ are equivalent to those usually used in mesoscopic wave physics \cite{vanrossum99,akkermans07}, except that they are defined for a given outgoing mode (mode $b$) with summations in their definitions running over all incoming modes, whereas the usual definition is for a given incoming mode with summations running over all outgoing modes. For a reciprocal medium, to which we will limit our consideration here, these two definitions are of course equivalent. We also note the appearance of a new quantity $V_b(t)$ in Eq.\ (\ref{eq:ConservationLaw}). These quantity vanishes in the absence of amplification, when $T_{b}\left(t\right)+R_{b}\left(t\right) = 1$. In an amplifying medium, its appearance in Eq.\ (\ref{eq:ConservationLaw}) illustrates that amplification [i.e. the condition $T_{b}\left(t\right)+R_{b}\left(t\right) > 1$] necessarily implies additional noise [the last sum in Eq.\ (\ref{eq:GeneralIOrelationTime})]. This can be interpreted as a manifestation of the fluctuation-dissipation theorem in this particular case.

\section{Average number of photocounts}

We now consider an ideal photodetector illuminated by the outgoing mode $b$. The operator corresponding to the number of photocounts registered by the photodetector during a sampling time $\tau$ is
\begin{equation}
\hat{n}_{b} = \int_{0}^{\tau} dt\, \hat{a}_{b}^{\dagger}(t)\hat{a}_{b}(t)
\label{eq:NbDefinition}
\end{equation}
Using Eqs.\ (\ref{eq:GeneralIOrelationTime}), (\ref{eq:ExpectationTime}) and the fact that all incoming modes except a single mode $a$ are in vacuum states, one readily calculates the quantum mechanical expectation value of ${\hat n}_b$:
\begin{equation}
\langle \hat{n}_{b} \rangle =\int_{0}^{\tau} dt\; T_{ab}\left(t\right)\langle \hat{a}_{a}^{\dagger}\left(t\right)\hat{a}_{a}\left(t\right)
\rangle - 
\eta(\omega_0) \frac{\Delta \omega}{2\pi}
\int_{0}^{\tau} dt\; V_{b}\left(t\right)
\label{eq:Nb}
\end{equation}
We will use angular brackets $\langle \ldots \rangle$ to denote quantum mechanical expectation values from here on. 

Equation (\ref{eq:Nb}) was obtained for arbitrary but given time-dependent functions $T_{ab}(t)$ and $V_b(t)$. $\langle {\hat n}_b \rangle$ can fluctuate from one realization of these two random functions to another. We can perform ensemble averaging, denoted by a horizontal bar $\overline{(\ldots)}$, over all possible random realizations of $T_{ab}(t)$ and $V_b(t)$:
\begin{equation}
\overline{\langle \hat{n}_{b} \rangle} =
\overline{T}_{ab} \langle  {\hat n}_a \rangle -
\frac{\tau \Delta \omega}{2\pi}
\eta(\omega_0) \overline{V}_{b}
\label{eq:Nba}
\end{equation}
where $\langle {\hat n}_a \rangle$ is the average number of photocounts that would be counted by the same photodetector in the incident mode $a$ during the same time $\tau$, and the random processes $T_{ab}(t)$ and $V_b(t)$ are assumed stationary, so that their statistical moments are independent of time.

\begin{figure}
\includegraphics[width=\columnwidth]{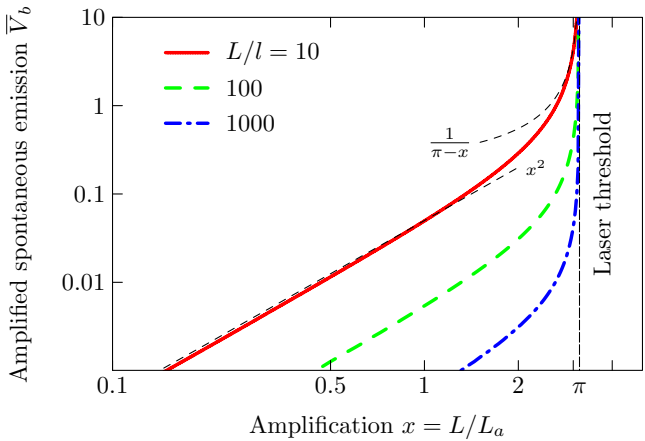}
\caption{\label{fig:Vb} Average amplified spontaneous emission coefficient $\overline{V}_{b}$ as a function of the ratio of the slab thickness $L$ to the amplification length $L_a$ for 3 different thicknesses $L/\ell = 10$, 100 and 1000. Dashed lines show asymptotes for $x \rightarrow 0$ and $x \rightarrow \pi$ in the case of $L/\ell = 10$.}
\end{figure}

Equation (\ref{eq:Nba}) contains two terms. The first one describes transmission of the incident light through the slab of amplifying medium, whereas the second one represents the amplified spontaneous emission (ASE) of the random medium. The average transmission and spontaneous emission coefficients $\overline{T}_{ab}$ and $\overline{V}_b$ entering Eq.\ (\ref{eq:Nba}) can be calculated in the diffusion approximation (see Appendix \ref{appav}):
\begin{eqnarray}
\overline{T}_{ab} &=& \frac{1}{N} \frac{\sin(\ell/L_a)}{\sin(L/L_a)}
\label{tab}
\\
\overline{V}_b &=& 
\frac{\sin(\ell/L_a) + \sin[(L-\ell)/L_a]}{\sin(L/L_a)} - 1
\label{vb}
\end{eqnarray}
where $L_a = \sqrt{\ell \ell_a/3}$ is the macroscopic amplification length and $\ell_a$ is the inverse amplification coefficient. Both $\overline{T}_{ab}$ and $\overline{V}_b$ diverge at $L/L_a = \pi$ which we identify as a laser threshold, following previous literature \cite{patra99b, patra99, zyuzin95, burkov97}. Obviously, our approach based on the linear equations (\ref{eq:GeneralIOrelationTime}) can only be applied below the threshold, i.e. for $L/L_a < \pi$, which is a condition for having a random laser amplifier and not a random laser. The average transmission coefficient of a random laser amplifier $\overline{T}_{ab}$ was previously studied in Refs.\ \cite{zyuzin95,burkov97}. However, the spontaneous emission coefficient $V_b$ was not introduced previously, even though ASE of a random amplifying medium was studied in Refs.\ \cite{bee98,patra99}. We show $\overline{V}_b$ as a function of $L/L_a$ in Fig.\ \ref{fig:Vb}.
The fraction of photocounts due to amplified spontaneous emission in the total signal can be found by dividing the second term in Eq.\ (\ref{eq:Nba}) by $\overline{\langle {\hat n}_b \rangle}$:
\begin{eqnarray}
\varphi = \frac{-\eta(\omega_0) \overline{V}_b \tau \Delta \omega/(2\pi)}{\overline{\langle {\hat n}_b \rangle}}
\label{fraction}
\end{eqnarray}
We will find this quantity useful later. It varies between 0 (no spontaneous emission) and 1 (no incoming light).

We will see that in the following an important role will be played by the average dimensionless conductance of the disordered slab in the absence of amplification ($L_a \rightarrow \infty$), equal to the transmittance $T$ and defined as
\begin{eqnarray}
g \equiv T = \frac{4}{3} \sum\limits_{\alpha, b} \overline{T}_{\alpha b} =
\frac{4}{3} N \frac{\ell}{L}
\label{g}
\end{eqnarray}
Because we do not treat the dependence of $\overline{T}_{\alpha b}$ on $\alpha$ and $b$ in full (in reality, $\overline{T}_{\alpha b}$ is not independent of $\alpha$ and $b$ but contains slowly varying dependences on the directions of propagation of corresponding modes \cite{vanrossum99}), we added a factor $\frac43$ in front of the sum over modes in Eq.\ (\ref{g}). This makes the final expression for $g$ consistent with its standard definition \cite{vanrossum99,akkermans07}. A transition from diffuse multiple scattering to Anderson localization is expected at $g \approx 1$ \cite{abrahams79}. In the present paper we only consider the case of $g \gg 1$.

\section{Variance of photocount distribution}
\label{var}

We now proceed to the calculation of the fluctuations in the photocount number. These fluctuations can be of classical or quantum origin. Classical fluctuations arise because the transmission coefficient $T_{ab}(t)$ fluctuates in time and from one realization of disorder to another. The fluctuations of $T_{ab}(t)$ can be characterized by its autocorrelation function
\begin{equation}
C_{T_{ab}T_{ab}}(t) = \frac{\overline{T_{ab}\left(t'\right)T_{ab}\left(t'+t\right)}}{\overline{T}_{ab}^{2}}-1
\label{eq:CTabTab}
\end{equation}
Quantum fluctuations appear due to the nonclassical nature of light in the incoming mode $a$ and spontaneous emission of light by the atoms in the random medium. The degree of ``quantumness'' of the incident light can be quantified by the Mandel parameter $Q_{a} = \mathrm{var}(\hat{n}_{a})/\langle \hat{n}_{a} \rangle - 1$ \cite{mandel95}, where
$\mathrm{var} ({\hat n}_a) = \langle {\hat n}_a^2 \rangle  - \langle {\hat n}_a \rangle^2$, whereas the intensity of spontaneous emission is controlled by the Bose-Einstein function $\eta(\omega_0)$. Because both the incident light and the spontaneous emission have to cross the whole or a part of the disordered medium in order to be detected in the outgoing mode $b$, quantum fluctuations will be ``weighted'' by the fluctuations of classical coefficients $T_{ab}$ and $V_b$ in the final result for the variance of ${\hat n}_b$. In particular, we will see that, in addition to (\ref{eq:CTabTab}), the following correlation functions appear:
\begin{eqnarray}
C_{T_{ab}V_{b}}(t) &=& \frac{\overline{T_{ab}\left(t'\right)V_{b}\left(t'+t\right)}}{\overline{T}_{ab} \overline{V}_{b}}-1
\label{eq:CTabVb}
\\
C_{V_{b}V_{b}}(t) &=& \frac{\overline{V_{b}\left(t'\right)V_{b}\left(t'+t\right)}}{\overline{V}_{b}^2}-1
\label{eq:CVbVb}
\end{eqnarray}
The calculation of correlation functions $C_{T_{ab}T_{ab}}(t)$, $C_{T_{ab}V_{b}}(t)$ and $C_{V_{b}V_{b}}(t)$ is presented in the Appendices \ref{appctabtab}, \ref{appctabvb} and \ref{appcvbvb}, respectively. In order to calculate these correlation functions, we excluded the immediate vicinity of the laser threshold from our consideration and assumed that $\Delta = 1 - L/\pi L_a \gg 1/\sqrt{g}$, where, we remind, $g \gg 1$.
 
With these definitions in hand, we start by calculating the quantum mechanical expectation value of ${\hat n}_b^2$,
\begin{eqnarray}
\langle \hat{n}_{b}^{2} \rangle = \int_{0}^{\tau}dt_{1}\int_{0}^{\tau}dt_{2}\langle \hat{a}_{b}^{\dagger}\left(t_{1}\right)\hat{a}_{b}
\left(t_{1}\right)\hat{a}_{b}^{\dagger}\left(t_{2}\right)
\hat{a}_{b}\left(t_{2}\right)\rangle  
\hspace{5mm}
\label{nb2}
\end{eqnarray}
In addition to the assumption of slow variations of $t_{\alpha b}(t)$, $r_{\beta b}(t)$ and $v_{\gamma b}^*(t)$ on the scale of $1/\Delta \omega$, adopted in the previous sections, we restrict our consideration to stationary, monochromatic incident light in the mode $a$, for which $I_a = \langle {\hat a}_a^{\dagger}(t) {\hat a}_a(t) \rangle$ and
$F_a = \langle {\hat a}_a^{\dagger}(t) {\hat a}_a^{\dagger}(t') {\hat a}_a(t) {\hat a}_a(t') \rangle$ are time-independent and
$\langle {\hat a}_a^{\dagger}(\omega) {\hat a}_a(\omega') \rangle = 2\pi I_a \delta(\omega-\omega') \delta(\omega-\omega_0)$.    
A relatively simple result for $\langle \hat{n}_{b}^{2} \rangle$ can be then obtained in the limit of long measurement time $\tau \gg 1/\Delta \omega$:
\begin{eqnarray}
\langle \hat{n}_{b}^{2} \rangle &=&
\langle {\hat n}_b \rangle +
F_a \int_0^{\tau} dt \int_0^{\tau} dt' T_{ab}(t) T_{ab}(t')
\nonumber \\
&-& 2 I_a \eta(\omega_0) \left[
\frac{\Delta \omega}{2\pi} \int_0^{\tau} dt \int_0^{\tau} dt' T_{ab}(t) V_{b}(t')
\right.
\nonumber \\
&+& \left. \int_0^{\tau} dt T_{ab}(t) V_{b}(t)
\right]
\nonumber \\
&+& \eta^2(\omega_0) \left\{
\left[ \frac{\Delta \omega}{2\pi} \int_0^{\tau} dt V_{b}(t)
\right]^2
\right.
\nonumber \\
&+& \left. \frac{\Delta \omega}{2\pi} \int_0^{\tau} dt V_{b}^2(t) \right\}
\label{nb2qe}
\end{eqnarray}
We will characterize fluctuations of ${\hat n}_b$ by the their variance $\mathrm{var} ({\hat n}_b) = \overline{\langle {\hat n}_b^2 \rangle} - (\overline{\langle {\hat n}_b \rangle})^2$, divided by the square of the average number of photocounts: $\delta_b^2 = \mathrm{var} ({\hat n}_b)/(\overline{\langle {\hat n}_b \rangle})^2$. To calculate $\delta_b^2$, we average Eq.\ (\ref{nb2qe}) over an ensemble of realizations of the disordered medium, use Eq.\ (\ref{eq:Nba}) for $\overline{\langle {\hat n}_b \rangle}$ and definition (\ref{fraction}) for $\varphi$. This yields
\begin{eqnarray}
\delta_b^2 &=&
\frac{1}{\overline{\langle {\hat n}_b \rangle}}
\left\{ 1 + \varphi \overline{T}_{b}
\left( \frac{2 \pi I_a}{N \Delta \omega} \right) \left[
\vphantom{\frac{\varphi}{1-\varphi}}
2 \left( 1 + C_{T_{ab} V_b}(0) \right)
\right. \right.
\nonumber \\
&+& \left. \left. \frac{\varphi}{1-\varphi} \left( 1 + C_{V_b V_b}(0) \right) \right] \right\}
\nonumber \\
&+& (1 - \varphi)^2 \left[ \left(1 + \frac{Q_a}{\langle {\hat n}_a \rangle} \right) \delta_{T_{ab} T_{ab}}^2 + \frac{Q_a}{\langle {\hat n}_a \rangle} \right]
\nonumber \\
&+& 2 \varphi (1 - \varphi) \delta_{T_{ab} V_{b}}^2 +
\varphi^2 \delta_{V_{b} V_{b}}^2
\label{db2}
\end{eqnarray}
where
$\delta_{T_{ab} T_{ab}}^{2} = (2/\tau)\int_{0}^{\tau} (1-t/\tau) C_{T_{ab}T_{ab}}(t)dt$ and similarly for 
$\delta_{T_{ab} V_{b}}^{2}$ and $\delta_{V_{b} V_{b}}^{2}$.
This equation is the main result of the paper.

To discuss the meaning of different terms in Eq.\ (\ref{db2}), let us first consider the situation when the spontaneous emission is absent ($\varphi = 0$). Eq.\ (\ref{db2}) then reduces to the result obtained previously in Ref.\ \cite{skip07}. It contains the shot noise $1/\overline{\langle {\hat n}_b \rangle}$ due to the fact that the energy of the electromagnetic field in the outgoing mode $b$ can be absorbed by the detector only in discrete portions (quanta) $\hbar \omega$, the classical noise $\delta_{T_{ab} T_{ab}}^2$ due to the fluctuations of the classical transmission coefficient $T_{ab}$, and a combined term $Q_a (1 + \delta_{T_{ab} T_{ab}}^2)/\langle {\hat n}_a \rangle$ that vanishes for light in the coherent (``classical'') state for which $Q_a = 0$, and can be either positive or negative for nonclassical light (e.g., light in Fock or squeezed states) \cite{skip07}. 
When the spontaneous emission is present ($\varphi >0$), the situation is more complex and all terms of Eq.\ (\ref{db2}) have to be taken into account. The terms of Eq.\ (\ref{db2}) containing $C_{V_b V_b}(0)$ and $\delta_{V_b V_b}^2$ originate from the spontaneous emission alone, whereas those containing $C_{T_{ab} V_b}(0)$ and $\delta_{T_{ab} V_b}^2$ represent the interference of the incoming light with the spontaneous emission. 

Because Eq.\ (\ref{db2}) is quite lengthy and difficult to analyze in its general form, we will restrict our consideration to two particularly interesting cases. The first case corresponds to a strong incoming wave such that $I_a \ne 0$ and the fraction of spontaneous emission in the total number of photocounts in the outgoing mode $b$ is small: $\varphi \ll 1$. In this limit, we will expand Eq.\ (\ref{db2}) in power series in $\varphi$ and will keep only linear terms (Section \ref{phismall}). The second case corresponds to the absence of incoming wave ($I_a = 0$ and $\varphi = 1$). The outgoing light in the mode $b$ is then due uniquely to the amplified spontaneous emission (Section \ref{phi1}).

\subsection{Strong incident wave}
\label{phismall}

\begin{figure}
\includegraphics[width=\columnwidth]{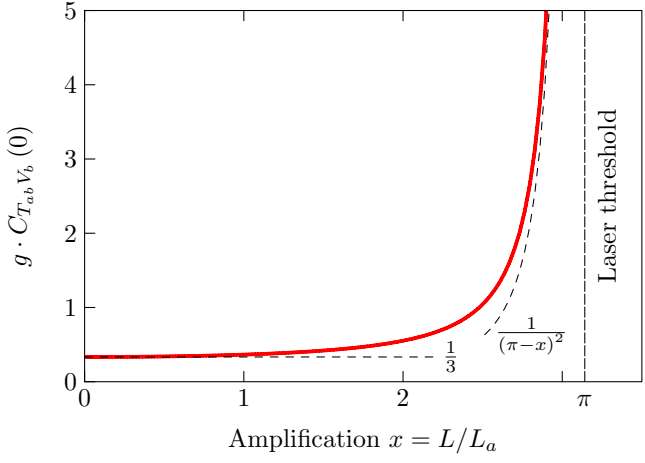}
\caption{\label{fig:CTabVb0}
Correlation function
$C_{T_{ab}V_{b}}(0) = \overline{T_{ab} V_b}/(\overline{T}_{ab} \overline{V}_b)-1$ as a function of the ratio of slab thickness to the amplification length, $x = L/L_a$. Dashed lines show asymptotes for $x \rightarrow 0$ and $x \rightarrow \pi$.}
\end{figure}

When the incident wave in the incoming mode $a$ is strong and the fraction of photocounts due to the amplified spontaneous emission in the outgoing mode $b$ is small, we can develop Eq.\ (\ref{db2}) in series in $\varphi \ll 1$ and keep only linear terms. To be specific, we assume that the incident wave is in the coherent state ($Q_a = 0$). After a substitution of Eq.\ (\ref{fraction}) for $\varphi$, the normalized variance of photocount fluctuations can be written as
\begin{eqnarray}
\delta_b^2 &=& \frac{1}{\overline{\langle {\hat n}_b \rangle}}
\left\{
1 - 2\eta(\omega_0) \overline{V}_b \left[
1 +  C_{T_{ab} V_b}(0) \right] \right\}
\nonumber \\
&+& \frac{2 \eta(\omega_0) \overline{V}_b}{\overline{T}_{b}}
\left( \frac{ N \Delta \omega}{2\pi I_a} \right)
\left( \delta_{T_{ab} T_{ab}}^2 - \delta_{T_{ab} V_{b}}^2 \right)
\nonumber \\
&+& \delta_{T_{ab} T_{ab}}^2
\label{db2phismall}
\end{eqnarray}
For the amplifying medium with time-independent, static properties, $\delta_{T_{ab} T_{ab}}^2 = C_{T_{ab} T_{ab}}(0) \sim 1$ and $\delta_{T_{ab} V_{b}}^2 = C_{T_{ab} V_{b}}(0) \sim 1/g$. The latter quantity is shown in Fig.\ \ref{fig:CTabVb0} as a function of the ratio of slab thickness $L$ to the amplification length $L_a$. Equation (\ref{db2phismall}) then becomes
\begin{eqnarray}
\delta_b^2 &=& \frac{1}{\overline{\langle {\hat n}_b \rangle}}
\left\{
1 - 2\eta(\omega_0) \overline{V}_b \left[
1 +  C_{T_{ab} V_b}(0) \right] \right\}
\nonumber \\
&+& \frac{2 \eta(\omega_0) \overline{V}_b}{\overline{T}_{b}}
\left(\frac{ N \Delta \omega}{2\pi I_a} \right)
\left( C_{T_{ab} T_{ab}}(0) - C_{T_{ab} V_b}(0) \right)
\nonumber \\
&+& C_{T_{ab} T_{ab}}(0)
\label{db2phismall1}
\end{eqnarray}
This result differs from the result of Patra and Beenakker \cite{patra99} because averaging over realizations of disorder was performed in a different way in that work. A result consistent with Ref.\ \cite{patra99} is obtained by setting $C_{T_{ab} T_{ab}}(0) = 0$ and $C_{T_{ab} V_{b}}(0) = 0$: $\delta_b^2 = [1 - 2 \eta(\omega_0) \overline{V}_b]/\overline{\langle {\hat n}_b \rangle}$. Obviously, this corresponds to neglecting fluctuations of $T_{ab}$ and $V_b$ from one realization of disorder to another.

\begin{figure}
\includegraphics[width=\columnwidth]{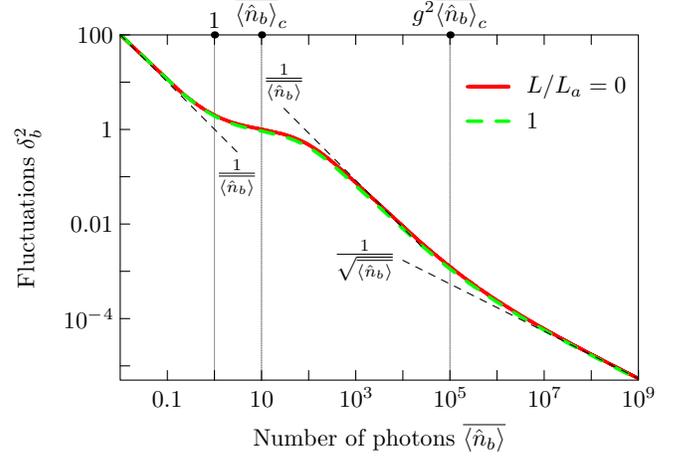}
\caption{\label{fig:SMdb}Normalized variance of photon number fluctuations in transmission of a strong incident wave through a slab of amplifying disordered medium.
For this figure, we set slab thickness $L/\ell = 100$, $\overline{\langle {\hat n}_b \rangle}_c = 10$,
$2\pi I_a/N \Delta \omega = 10$, $g = 100$ and $\eta(\omega_0) = -1$  (i.e. $T \rightarrow 0^{-}$). The two curves correspond to $L/L_a = 0$ (no amplification) and 1.
Note that the curve corresponding to $L/L_a = 1$ is slightly below the $L/L_a = 0$ curve.}
\end{figure}

Surprisingly, the different way of averaging over disorder may have an important influence on seemingly very general conclusions that one makes by analyzing the variance of ${\hat n}_b$. Indeed, in the limit of weak amplification $\overline{V}_b \ll 1$ and for $g \gg 1$ and $\tau \Delta \omega \gg 1$, we readily obtain from Eq.\ (\ref{db2phismall1}) that $\mathrm{var}\, {\hat n}_b = \delta_b^2 (\overline{\langle {\hat n}_b \rangle})^2 \simeq \overline{\langle {\hat n}_b \rangle} + (\overline{\langle {\hat n}_b \rangle})^2 + 2 \eta(\omega_0) \overline{V}_b \overline{\langle {\hat n}_b \rangle} (\tau \Delta \omega/2\pi)$, where the first term describes the shot noise, the second --- large intensity fluctuations due to disorder in the medium \cite{shapiro86}, and the last one can be interpreted as an excess noise due to the amplified spontaneous emission. Note that because $\eta(\omega_0) < 0$ in an amplifying medium, the excess noise is {\em negative}. This result, also illustrated in Fig.\ \ref{fig:SMdb}, is a consequence of the interplay between quantum fluctuations, due to the interference of the incident light with the amplified spontaneous emission, and classical fluctuations, due to the randomness of the transmission and spontaneous emission coefficients $T_{ab}$ and $V_b$.
It seems to be in conflict with the positive excess noise found in Ref.\ \cite{patra99}, but the solution of this contradiction lies in different ways of averaging over disorder: we consider $\mathrm{var}\, {\hat n}_b = \overline{\langle {\hat n}_b^2 \rangle} - (\overline{\langle {\hat n}_b \rangle})^2$ instead of  $\mathrm{var'} {\hat n}_b = \overline{\langle {\hat n}_b^2 \rangle - \langle {\hat n}_b \rangle^2}$ in Ref.\ \cite{patra99}. When we calculate the latter quantity, we find $\mathrm{var'} {\hat n}_b = \overline{\langle {\hat n}_b \rangle} - 2 \eta(\omega_0) \overline{V}_b \overline{\langle {\hat n}_b \rangle}$ and the excess noise is positive, in agreement with \cite{patra99}. An important conclusion of this analysis is that if theoretical results are compared to experiments, care should be taken to ensure that averaging is performed in an appropriate way. In a medium with fluctuating properties, the simultaneous quantum and disorder averages $\overline{\langle \ldots \rangle}$ that we use is the only experimentally feasible option because both are typically replaced by time averages and it is hence not possible to perform the quantum average $\langle \ldots \rangle$ without performing averaging over disorder $\overline{(\ldots)}$ \cite{balog06,skip07}.  

Let us now analyze the behavior of Eq.\ (\ref{db2phismall}) as a function of $\overline{\langle {\hat n}_b \rangle}$. To be more specific, we restrict our consideration to the case when the random medium represents itself an ensemble of point-like scattering centers in Brownian motion with diffusion coefficient $D_B$. In this situation, the motion of scatterers introduces a characteristic time $t_c = \frac23 t_0 (\ell/L)^2$, where $t_0 = 1/4 k^2 D_B$. This is the typical correlation time of transmitted light. We find convenient to introduce the average number of photons $\overline{\langle {\hat n}_b \rangle}_c$ detected during a sampling time equal to $t_c$. 
Three distinct regimes can then be identified in the dependence of  $\delta_b^2$ on $\overline{\langle {\hat n}_b \rangle}$. First, for $\overline{\langle {\hat n}_b \rangle} < 1$, Eq.\ (\ref{db2phismall}) is dominated by the first term on its right-hand side: $\delta_b^2 \sim 1/\overline{\langle {\hat n}_b \rangle}$.
Second, when $\overline{\langle {\hat n}_b \rangle}$ becomes of the order of 1, another two terms on the right-hand side of Eq.\ (\ref{db2phismall}) come into play. In the limit of $\overline{\langle {\hat n}_b \rangle} > \overline{\langle {\hat n}_b \rangle}_c$, we find from the results of Appendix \ref{appctabtab} that
$\delta_{T_{ab} T_{ab}}^{2} \propto C_1 \overline{\langle {\hat n}_b \rangle}_c/\overline{\langle {\hat n}_b \rangle} + (C_2/g) (\overline{\langle {\hat n}_b \rangle}_c/\overline{\langle {\hat n}_b \rangle})^{1/2}$, where $C_1$ and $C_2$ are constants and the two contributions to $\delta_{T_{ab} T_{ab}}^{2}$ are due to the correlation functions $C_{T_{ab} T_{ab}}^{(1)}(t)$ and $C_{T_{ab} T_{ab}}^{(2)}(t)$, respectively.
These are the short- and long-range correlation functions defined in Appendix \ref{appctabtab}.
Using results of Appendix \ref{appctabvb}, we also find $\delta_{T_{ab} V_{b}}^{2} \propto (1/g) (\overline{\langle {\hat n}_b \rangle}_c/\overline{\langle {\hat n}_b \rangle})^{1/2}$.
As long as $\overline{\langle {\hat n}_b \rangle} < g^2 \overline{\langle {\hat n}_b \rangle}_c$, the first term of $\delta_{T_{ab} T_{ab}}^{2}$ dominates and we obtain $\delta_b^2 \propto 1/\overline{\langle {\hat n}_b \rangle}$, similarly to the case of $\overline{\langle {\hat n}_b \rangle} < \overline{\langle {\hat n}_b \rangle_c}$, but with a different coefficient. However, the truly large-$\overline{\langle {\hat n}_b \rangle}$ behavior is reached for $\overline{\langle {\hat n}_b \rangle} > g^2 \overline{\langle {\hat n}_b \rangle}_c$, when the second term of $\delta_{T_{ab} T_{ab}}^{2}$ and $\delta_{T_{ab} V_{b}}^{2}$ become larger than the first term of $\delta_{T_{ab} T_{ab}}^{2}$; $\delta_b^2 \propto 1/\overline{\langle {\hat n}_b \rangle}^{1/2}$ in this limit. 
We illustrate the three regimes in the dependence of $\delta_b^2$ on $\overline{\langle {\hat n}_b \rangle}$ and transitions between them in Fig.\ \ref{fig:SMdb}.

\subsection{Noise in the amplified spontaneous emission}
\label{phi1}

\begin{figure}
\includegraphics[width=\columnwidth]{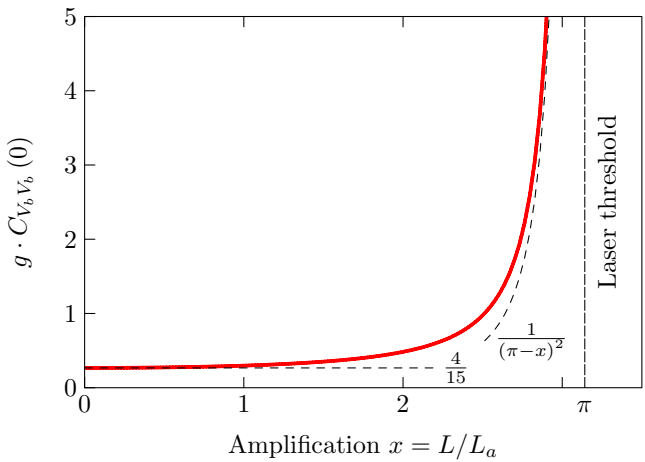}
\caption{\label{fig:CVbVb0}
Variance of fluctuations of the spontaneous emission coefficient $V_b$: $C_{V_{b}V_{b}}(0) = \overline{V_b^2}/\overline{V_b}^2 -1$ as a function of the ratio of slab thickness $L$ to the amplification length $L_a$.}
\end{figure}

\begin{figure}
\includegraphics[width=\columnwidth]{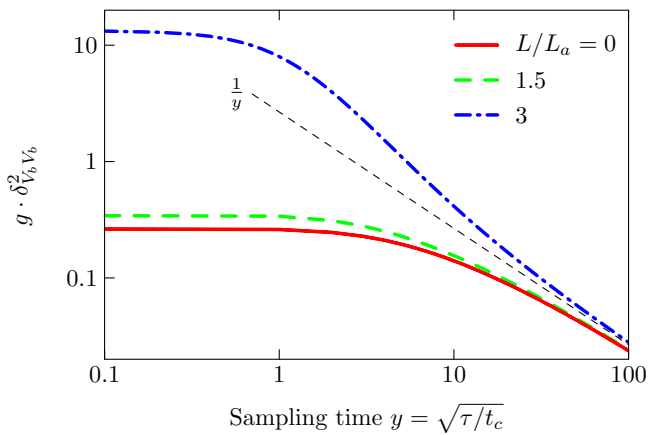}
\caption{\label{fig:dVbVb}
The most interesting contribution to the normalized variance of photocount fluctuations, $\delta_{V_{b}V_{b}}^{2}$, as
a function of the square root of normalized sampling time $y = \sqrt{\tau/t_c}$ for several values of the ratio $L/L_a$.}
\end{figure}

We saw in the previous section that the interference of a strong incident wave in the coherent state with the weak amplified spontaneous emission in an amplifying random medium modifies noise properties of the transmitted light. We now consider another special case of the general Eq.\ (\ref{db2}): a situation when no light is sent into the random medium from the outside (i.e. $\langle {\hat n}_a \rangle = 0$) and the signal measured in the outgoing mode $b$ is due uniquely to the amplified spontaneous emission of the medium itself. One readily sees from Eqs.\ (\ref{eq:Nba}) and (\ref{db2}) that in this case
\begin{eqnarray}
\overline{\langle \hat{n}_{b} \rangle} &=&
-\frac{\tau\Delta\omega}{2\pi}
\eta(\omega_0) \overline{V}_b
\label{ase1}
\\
\delta_b^2 &=&
\frac{1}{\overline{\langle \hat{n}_b \rangle}}
\left[ 1 - \eta(\omega_0) \overline{V}_b
\left( 1 + C_{V_b V_b}(0) \right) \right] + \delta_{V_b V_b}^2
\hspace{10mm}
\label{ase2}
\end{eqnarray}
Once again, we recover the result obtained previously \cite{bee98},
$\delta_b^2 = 1/\overline{\langle \hat{n}_b \rangle} + 2\pi/\tau \Delta \omega$, if we neglect fluctuations of $V_b$ from one realization of disorder to another and set $C_{V_b V_b}(0) = 0$ and $\delta_{V_b V_b} = 0$. It is true that these fluctuations are weak: $C_{V_b V_b}(0) \sim 1/g \ll 1$ and $\delta_{V_b V_b} \sim 1/g \ll 1$ in the limit of weak amplification, but they grow when the laser threshold is approached and even diverge at the threshold (see Figs.\ \ref{fig:CVbVb0} and \ref{fig:dVbVb}). 

\begin{figure}
\includegraphics[width=\columnwidth]{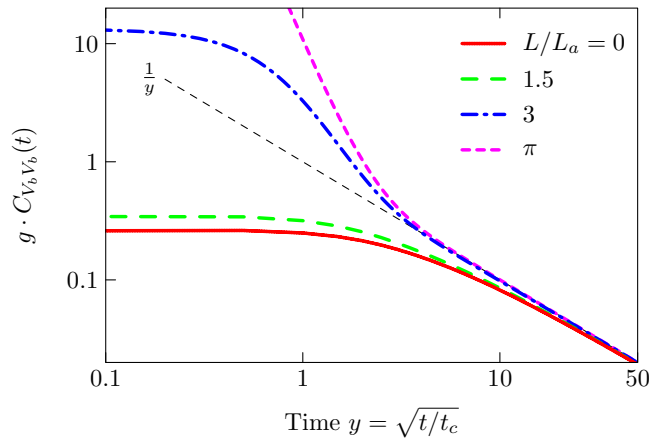}
\caption{\label{fig:CVbVb}
Correlation function $C_{V_{b}V_{b}}(t)$ as a function of the square root of normalized time $y = \sqrt{t/t_c}$ for several values of the ratio $L/L_a$.}
\end{figure}

In order to study the behavior of Eq.\ (\ref{ase2}) as a function of $\overline{\langle {\hat n}_b \rangle}$, we first consider how $\delta_{V_b V_b}^2$ depends on the sampling time $\tau$. $\delta_{V_b V_b}^2$ is an integral of $C_{V_b V_b}(t)$ multiplied by $(1-t/\tau)$ and divided by $\tau/2$. For an ensemble of scattering centers in Brownian motion, $C_{V_b V_b}(t)$ is calculated in the Appendix \ref{appcvbvb} and we show it in Fig.\ \ref{fig:CVbVb} for several values of $L/L_a$. In the limit of short sampling times $\tau$, we have $\delta_{V_b V_b}^2 = C_{V_b V_b}(0)$ with $C_{V_b V_b}(0)$ given by Eq.\ (\ref{rescvbvb0}) of Appendix \ref{appcvbvb} and shown as a function of $L/L_a$ in Fig.\ \ref{fig:CVbVb0}. In the limit of long sampling times, the scaling of $\delta_{V_b V_b}^2$  with $\tau$ is determined by the long-time asymptotics of $C_{V_b V_b}(t)$. It follows from Eq.\ (\ref{rescvbvb}) of Appendix \ref{appcvbvb} that $C_{V_b V_b}(t) = \sqrt{t_c/t}/g$ with $t_c = \frac23 t_0(\ell/L)^2$ in the limit of $t \gg t_c$. This yields $\delta_{V_b V_b}^2 = (8/3g) \sqrt{t_c/\tau}$. This asymptotic behavior of $\delta_{V_b V_b}^2$ is illustrated in Fig.\ \ref{fig:dVbVb}.
 Because according to Eq.\ (\ref{ase1}), $\overline{\langle {\hat n}_b \rangle} \propto \tau$, we can replace the ratio $t_c/\tau$ by a ratio $\overline{\langle {\hat n}_b \rangle}_c/\overline{\langle {\hat n}_b \rangle}$ with $\overline{\langle {\hat n}_b \rangle}_c$ denoting the average number of photocounts in a sampling time equal to $t_c$. We therefore obtain
$\delta_{V_b V_b}^2 = (8/3g) \sqrt{\overline{\langle {\hat n}_b \rangle}_c/\overline{\langle {\hat n}_b \rangle}}$.

\begin{figure}
\includegraphics[width=\columnwidth]{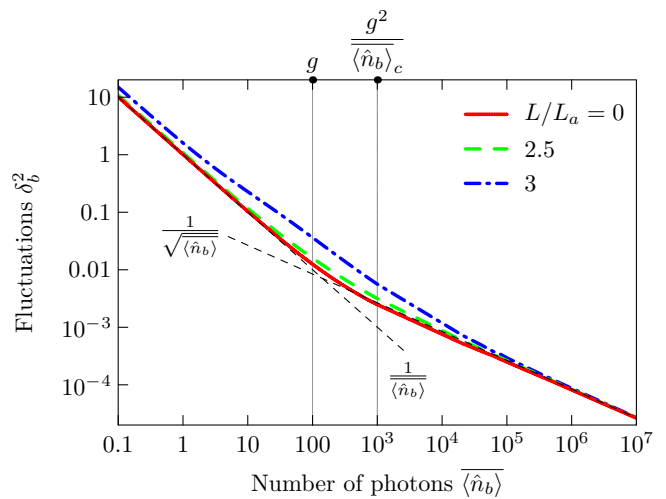}
\caption{\label{fig:NRdb}
Normalized variance of photocount fluctuations 
$\delta_{b}^{2}$ resulting from the amplified spontaneous emission of a disordered slab of thickness $L$. The three curves correspond to three different values of the ratio of $L$ to the amplification length $L_a$. Other parameters are set to $L/\ell = 100$, $\overline{\langle {\hat n}_b \rangle}_c = 10$, $g = 100$ and $\eta(\omega_0) = -1$.}
\end{figure}

We are now in a position to analyze the behavior of $\delta_b^2$ as a function of $\overline{\langle {\hat n}_b \rangle}$. At small $\overline{\langle {\hat n}_b \rangle}$, $\delta_{V_b V_b}^2$ can be neglected with respect to the first term in Eq.\ (\ref{ase2}) and we obtain $\delta_b^2 \propto 1/\overline{\langle {\hat n}_b \rangle}$. But even though initially small, $\delta_{V_b V_b}^2$ decays with $\overline{\langle {\hat n}_b \rangle}$ only as $1/\overline{\langle {\hat n}_b \rangle}^{1/2}$, which is due to the long-range nature of the correlation function $C_{V_b V_b}(t)$. As a consequence, $\delta_{V_b V_b}^2$ dominates the results in the limit of large $\overline{\langle {\hat n}_b \rangle}$:
\begin{eqnarray}
\delta_b^2 = \frac{8}{3 g}\sqrt{\frac{\overline{\langle {\hat n}_b \rangle}_c}{\overline{\langle {\hat n}_b \rangle}}},
\;\;\; \overline{\langle {\hat n}_b \rangle} \rightarrow \infty
\label{ase3}
\end{eqnarray}
The transition between the small- and large-$\overline{\langle {\hat n}_b \rangle}$ regimes in the dependence of $\delta_b^2$ on $\overline{\langle {\hat n}_b \rangle}$ takes place when $1/\overline{\langle {\hat n}_b \rangle} \sim \delta_{V_b V_b}^2$, i.e. at $\overline{\langle {\hat n}_b \rangle} \sim g$ if $\overline{\langle {\hat n}_b \rangle}_c > g$ or at $\overline{\langle {\hat n}_b \rangle} \sim g^2/ \overline{\langle {\hat n}_b \rangle}_c$ otherwise. We illustrate the transition between the two regimes in Fig.\ \ref{fig:NRdb}.  

Our result (\ref{ase3}) calls for two important comments. First, it makes an interesting link between random lasers and Anderson localization \cite{anderson58}. Indeed, Eq.\ (\ref{ase3}) states that in the limit of large average photon numbers, the noise of the amplified spontaneous emission is proportional to $1/g$, with $g$ --- the dimensionless conductance of the disordered medium --- determining the closeness of the Anderson localization transition, expected at $g = 1$ \cite{abrahams79}. This suggests that $g$ of a disordered sample can be measured by observing fluctuations of its amplified spontaneous emission. Another interesting observation is that Eq.\ (\ref{ase3}) contains the average number of photocounts in a time interval equal to the correlation time of light scattered in the disordered medium, $t_c$. Therefore, it appears possible to obtain useful information about the dynamics of scattering centers in a random medium (e.g., the diffusion coefficient of particles in suspension) from the fluctuations of amplified spontaneous emission. This ``noise spectroscopy'' can be made even more efficient by using the full equation (\ref{ase2}) to fit the whole curve $\delta_b^2(\overline{\langle {\hat n}_b \rangle})$ and not only its large-$\overline{\langle {\hat n}_b \rangle}$ part.

\section{Photon correlation spectroscopy in a random laser amplifier}
\label{corr}

In addition to the variance, another important quantity characterizing a fluctuating signal is the autocorrelation function of its fluctuations. The technique that uses the autocorrelation function of photon number fluctuations to characterize disordered media is known as photon correlation spectroscopy (PCS) \cite{weitz93}. Typically, one measures the autocorrelation function of photocounts detected by a photodetector in a mode $b$ of light scattered by a disordered sample:
\begin{equation}
C_{n_{b}n_{b}}(t) = \frac{\overline{\langle \hat{n}_{b}(t') \hat{n}_{b}(t'+t)\rangle }}{\overline{\langle \hat{n}_{b}(t')\rangle }\cdot\overline{\langle \hat{n}_{b}(t'+t)\rangle}}-1
\label{eq:CNbNb}
\end{equation}
where
\begin{equation}
\hat{n}_{b}(t') = \int_{t'}^{t'+\tau} dt''  \hat{a}_{b}^{\dagger}\left(t''\right)\hat{a}_{b}\left(t''\right)
\label{eq:NbDefinition2}
\end{equation}
is the operator corresponding to the number of photocounts measured by the photodetector in the time interval $[t', t' + \tau]$.
In a medium with fluctuating properties (like, e.g., a suspension of dielectric particles), $C_{n_{b}n_{b}}(t)$ contains information about the type and the intensity of fluctuations intrinsic to the medium (like, e.g., the type of particle motion --- Brownian, directed, arrested dynamics, etc., --- and its typical velocity). In optically dense media where light undergoes multiple scattering, PCS is most often called diffusing-wave spectroscopy (DWS) \cite{weitz93,maret97}.

The autocorrelation function (\ref{eq:CNbNb}) of light transmitted through a slab of amplifying random medium can be calculated using methods developed in Section \ref{var}. In fact, the normalized variance of photon number fluctuations $\delta_b^2$ is nothing else than $C_{n_{b}n_{b}}(0)$. In the present section, we will restrict our consideration to $t > \tau$, i.e. we will be interested in the correlation of photocounts corresponding to two nonoverlapping time intervals of equal duration $\tau$. Using Eq.\ (\ref{eq:GeneralIOrelationTime}), commutation relations (\ref{commat}), (\ref{eq:CommutationsTime}) and Eq.\ (\ref{eq:ExpectationTime}), we readily obtain
\begin{eqnarray}
&&\langle \hat{n}_{b}(t') \hat{n}_{b}(t'+t) \rangle = \int\limits_{t'}^{t'+\tau} dt_{1} \int\limits_{t'+t}^{t'+t+\tau} dt_2
\nonumber \\
&&\left[\vphantom{\left(\frac{\Delta\omega}{2\pi}\right)^{2}}
T_{ab}(t_{1}) T_{ab}(t_{2})
\langle \hat{a}_{a}^{\dagger}(t_1) \hat{a}_{a}^{\dagger}(t_2) \hat{a}_{a}(t_1) \hat{a}_{a}(t_2)\rangle \right.
\nonumber \\
&&\hspace{5mm} - \left. 2\frac{\Delta\omega}{2\pi} \eta(\omega_{0})
T_{ab}(t_{1}) V_b(t_2)
\langle \hat{a}_{a}^{\dagger}(t_{1}) \hat{a}_{a}(t_{1})\rangle
\right.
\nonumber \\
&&\hspace{5mm} + \left. \left(\frac{\Delta\omega}{2\pi}\right)^{2} \eta^{2}(\omega_{0})
V_{b}(t_{1})V_{b}(t_{2}) \right]
\label{eq:<Nb(t1)Nb(t2)>}
\end{eqnarray}
where the limit of long sampling times $\tau \gg 1/\Delta \omega$ was taken.
Note that here, in contrast to Eq.\ (\ref{nb2qe}) of Section \ref{var}, we did not assume that $\langle \hat{a}_{a}^{\dagger}(t_{1}) \hat{a}_{a}(t_{1})\rangle$ and $\langle \hat{a}_{a}^{\dagger}(t_1) \hat{a}_{a}^{\dagger}(t_2) \hat{a}_{a}(t_1) \hat{a}_{a}(t_2)\rangle$ are time-independent. We will not need this assumption in this section.

We now average Eq.\ (\ref{eq:<Nb(t1)Nb(t2)>}) over realizations of disorder and assume that the correlation functions $C_{T_{ab} T_{ab}}(t)$, $C_{T_{ab} V_{b}}(t)$ and $C_{V_{b} V_{b}}(t)$ do not vary significantly on the time scale of $\tau$. (Please note that this situation is different from the one considered in Section \ref{var}, where the main idea was precisely to take into account fluctuations of $T_{ab}$ and $V_b$ on times {\em shorter\/} than $\tau$.)
After a series of algebraic transformations we finally obtain
\begin{eqnarray}
C_{n_b n_b}(t) &=&
(1-\varphi)^2 \left\{
\left[1 + C_{n_a n_a}(t) \right] C_{T_{ab} T_{ab}}(t)
\right. \nonumber \\
&+& \left. C_{n_a n_a}(t) \right\}
+ 2 \varphi (1-\varphi) C_{T_{ab} V_{b}}(t)
\nonumber \\ 
&+& \varphi^2 C_{V_{b} V_{b}}(t)  
\label{cnn}
\end{eqnarray}
where $\varphi$ is the fraction of photocounts due to the amplified spontaneous emission defined by Eq.\ (\ref{fraction}), the correlation function of photocounts in the incident mode $a$, $C_{n_a n_a}(t)$, is defined in the same way as $C_{n_b n_b}(t)$, and $C_{T_{ab} T_{ab}}(t)$, $C_{T_{ab} V_{b}}(t)$ and $C_{V_{b} V_{b}}(t)$ are defined by Eqs.\ (\ref{eq:CTabTab})--(\ref{eq:CVbVb}).

The structure of Eq.\ (\ref{cnn}) is similar to that of Eq.\ (\ref{db2}). The first term is the only to survive in the absence of amplified spontaneous emission ($\varphi = 0$). In this case, Eq.\ (\ref{cnn}) reduces to the result previously obtained in Ref.\ \cite{skip07}. The second term, proportional to $\varphi (1-\varphi)$, represents the interference of external light with the amplified spontaneous emission and requires that both are present (i.e. that $0 < \varphi < 1$). Expanding Eq.\ (\ref{cnn}) in series in $\varphi \ll 1$, we can see that, similarly, to what happens to $\delta_b^2$, the amplified spontaneous emission provides a small correction to $\varphi = 0$ result. 
The most interesting limit of Eq.\ (\ref{cnn}) is the limit of $\varphi = 1$, corresponding to the absence of external illumination. In this limit, only the last term of Eq.\ (\ref{cnn}), representing the autocorrelation function of the amplified spontaneous emission, survives and
\begin{eqnarray}
C_{n_b n_b}(t) &=& C_{V_{b} V_{b}}(t)  
\label{cnnase}
\end{eqnarray}
For an ensemble of point-like scattering centers in Brownian motion, this correlation function is given by Eq.\ (\ref{rescvbvb}) of Appendix \ref{appcvbvb} and we show it in Fig.\ \ref{fig:CVbVb}. For weak amplification $x = L/L_a \ll 1$, it is small in magnitude ($\sim 1/g$) for $g \gg 1$, but it grows with $x$. $C_{V_b V_b}(0)$ diverges at the laser threshold $x = \pi$ (see Fig.\ \ref{fig:CVbVb0}). Interestingly, $C_{V_{b} V_{b}}(t)$ is long-range in time and decays only as $1/\sqrt{t}$. Long-range {\em spatial\/} correlations of similar origin have been previously reported by Patra and Beenakker \cite{patra99b}. A new feature of our result (\ref{cnnase}) is that it contains information about the dynamics of fluctuations in the disordered medium (e.g., information about dynamics of scattering centers). Measurements of $C_{n_b n_b}(t)$ can therefore be used for spectroscopy of disordered media, even in the absence of external light source.

\section{Conclusions}
\label{concl}

We presented a theoretical study of noise in the number of photocounts measured by an ideal, fast photodetector illuminated by a single mode $b$ of light emerging from an amplifying disordered medium (random laser amplifier) with fluctuating properties. Assuming that the photodetector is sensitive only to light in a small frequency band $\Delta \omega$, we derived general expressions for the normalized variance $\delta_b^2$ and autocorrelation function $C_{n_b n_b}(t)$ of photocount fluctuations in the limit of long sampling times $\tau \gg 1/\Delta \omega$. Because light in the mode $b$ contains a mixture of transmitted incident light and light spontaneously emitted and then amplified by the medium, the fraction of light due to the amplified spontaneous emission $\varphi$ appears as a natural parameter that controls the behavior of $\delta_b^2$ and $C_{n_b n_b}(t)$.
For $\delta_b^2$, two distinct regimes were identified and studied in detail. First, when $\varphi \ll 1$, the overall behavior of $\delta_b^2$ remains qualitatively similar to that at $\varphi = 0$. For a suspension of scattering centers in Brownian motion, for example, we found $\delta_b^2 \propto 1/\overline{\langle {\hat n}_b \rangle}$ in the limit of $\overline{\langle {\hat n}_b \rangle} \rightarrow 0$ and $\delta_b^2 \propto 1/g \overline{\langle {\hat n}_b \rangle}^{1/2}$ in the limit of $\overline{\langle {\hat n}_b \rangle} \rightarrow \infty$, with $g$ the dimensionless conductance of the disordered sample in the absence of amplification.
Second, when $\varphi = 1$, i.e. in the absence of external illumination, our result describes statistical properties of light spontaneously emitted by a random laser amplifier. We have found $\delta_b^2 \propto 1/g \overline{\langle {\hat n}_b \rangle}^{1/2}$ in the limit of $\overline{\langle {\hat n}_b \rangle} \rightarrow \infty$ for this case. 

For $C_{n_b n_b}(t)$, the two regimes $\varphi \ll 1$ and $\varphi = 1$ exist as well, with the most interesting result $C_{n_b n_b}(t) \propto 1/g \sqrt{t}$ obtained for the latter one. This long-range time correlation of the amplified spontaneous emission could possibly be used to probe dynamics of scattering centers in random media.

Our results allow us to establish an interesting link between random amplifying media and Anderson localization. We have found that for the amplified spontaneous emission, both $\delta_b^2$ and $C_{n_b n_b}(t)$ are inversely proportional to the dimensionless conductance $g$ of the disordered medium (we remind that the localization transition is expected at $g = 1$). Our results suggest that measurements of $\delta_b^2$ and $C_{n_b n_b}(t)$ in a random laser amplifier could allow a precise determination of $g$. 

\acknowledgements
This work is supported by the French ANR (project No. 06-BLAN-0096 CAROL) and the French Ministry of Education and Research (Research-Educational Network ``Quantum optics of random media'').

\appendix

\section{Average transmission, reflection and spontaneous emission coefficients of an amplifying disordered medium}
\label{appav}

In this Appendix, as well as in the three next Appendicies \ref{appctabtab}, \ref{appctabvb} and \ref{appcvbvb}, we will ignore the vector nature of electromagnetic waves, except for the expression of the number of transverse modes $N = k^2 A/2 \pi$ which is calculated with a proper account for two independent polarizations states.
Such a simplification is sufficient for our purposes because the key properties of average transmission, reflection and spontaneous emission coefficients, as well as the properties of their correlations functions, that we use in the main text, are the same for vector and scalar waves.

For distances exceeding the mean free path $\ell$, the transport of light in disordered media can be described in the diffusion approximation. A fundamental quantity through which many other quantities can be expressed is then the Green's function of the diffusion equation $G(\textbf{r}, \textbf{r}')$  obeying
\begin{eqnarray}
D \left( \nabla^2 - \chi^2 \right) G(\textbf{r}, \textbf{r}') &=& \delta(\textbf{r}-\textbf{r}')
\label{appdifeq}
\end{eqnarray} 
where $D = c\ell/3$ is the photon diffusion coefficient, $c$ is the speed of light in the medium, and $\chi$ is a parameter that we will specify later. In a slab of thickness $L$ and surface $A \gg L^2$, perpendicular to the $z$ axis, boundary conditions are $G = 0$ at the surfaces $z = 0$ and $z = L$, and the solution of Eq.\ (\ref{appdifeq}) reads
\begin{eqnarray}
&&G(\textbf{r}, \textbf{r}') = - \frac{1}{D (2 \pi)^2}
\int d^2 \textbf{q}
e^{-i \textbf{q} (\boldsymbol\rho - \boldsymbol\rho')}
\nonumber \\
&&\frac{\sinh\left[\sqrt{q^2 + \chi^2}z_{<}\right] \sinh\left[\sqrt{q^2 + \chi^2}(L-z_{>})\right]}{\sqrt{q^2 + \chi^2} \sinh\left(\sqrt{q^2 + \chi^2}L \right)}
\hspace{10mm} 
\label{appgreen}
\end{eqnarray}
with $\textbf{r} = (\boldsymbol\rho, z)$, $z_{<} = \mathrm{min}(z,z')$ and $z_{>} = \mathrm{max}(z,z')$.
The average intensity of light inside a slab of amplifying medium $\overline{I}_{\alpha}(\textbf{r})$, resulting from a plane wave incident on the slab in the mode $\alpha$, obeys the diffusion equation (\ref{appdifeq}) with $\chi^2 = -1/L_a^2$ and $-\delta(z - \ell)/A$ on the right-hand side. Here $L_a = \sqrt{\ell \ell_a/3}$ is the macroscopic amplification length and $1/\ell_a$ is the amplification coefficient. $\overline{I}_{\alpha}(\textbf{r})$ can be therefore expressed through the Green's function (\ref{appgreen}) as
\begin{eqnarray}
\overline{I}_{\alpha}(\textbf{r}) \simeq -\frac{1}{A} \int d^3 \textbf{r}' G(\textbf{r}, \textbf{r}') \delta(z'-\ell)
\label{appi}
\end{eqnarray}
with the integral over the volume of the slab. 
The average transmission and reflection coefficients from the incoming modes $\alpha$ and $\beta$, respectively, to the outgoing mode $b$ are
\begin{eqnarray}
\overline{T}_{\alpha b} &=& -\frac{4 \pi D}{k^2} \frac{\partial}{\partial z} \overline{I}_{\alpha}(\textbf{r})\left|_{z = L}\right.
= \frac{1}{N} \frac{\sin(\ell/L_a)}{\sin(L/L_a)}
\label{apptab}
\\
\overline{R}_{\beta b} &=& \frac{4 \pi D}{k^2} \frac{\partial}{\partial z} \overline{I}_{\beta}(\textbf{r})\left|_{z = 0}\right.
= \frac{1}{N} \frac{\sin[(L-\ell)/L_a]}{\sin(L/L_a)}
\hspace{10mm}
\label{apprab}
\end{eqnarray}
Within the approximations that we made in this paper, these results are independent of the mode indices $\alpha$, $\beta$ and $b$. More accurate considerations show that both $\overline{T}_{\alpha b}$ and $\overline{R}_{\beta b}$ exhibit slow dependences on $\alpha$, $\beta$ and $b$ \cite{vanrossum99,akkermans07}. In addition, a coherent backscattering cone exists in reflection \cite{akkermans07}. These corrections to  $\overline{T}_{\alpha b}$ and $\overline{R}_{\beta b}$ might be important if a quantitative comparison with experiments is attempted, but we neglect them here because they introduce additional technical complications in the analysis without bringing any qualitatively new physics.

The total transmission and reflections coefficients, as well as the spontaneous emission coefficient then follow from Eqs.\ (\ref{apptab}) and (\ref{apprab}):
\begin{eqnarray}
\overline{T}_{b} &=& \sum_{\alpha} \overline{T}_{\alpha b} =
\frac{\sin(\ell/L_a)}{\sin(L/L_a)}
\label{apptb}
\\
\overline{R}_{b} &=& \sum_{\beta} \overline{R}_{\beta b} =
\frac{\sin[(L-\ell)/L_a]}{\sin(L/L_a)}
\label{apprb}
\\
\overline{V}_{b} &=& \overline{T}_{b} + \overline{R}_{b} - 1
\nonumber \\
&=& \frac{\sin(\ell/L_a) + \sin[(L-\ell)/L_a]}{\sin(L/L_a)} - 1
\label{appvb}
\end{eqnarray} 
This completes derivation of Eqs.\ (\ref{tab}) and (\ref{vb}) of the main text.

\section{Correlation function of $T_{\alpha b}(t)$}
\label{appctabtab}

The correlation function $C_{T_{\alpha b} T_{\alpha b}}(t)$ defined in Eq.\ (\ref{eq:CTabTab}) can be decomposed into a sum of a large but short-range part $C_{T_{\alpha b} T_{\alpha b}}^{(1)}(t)$ and a small but long-range part $C_{T_{\alpha b} T_{\alpha b}}^{(2)}(t)$ \cite{vanrossum99, akkermans07, berk94,burkov97}. Because we will need it later, we consider a more general object $C_{T_{\alpha b} T_{\alpha' b}}(t)$ where $\alpha'$ can be different from $\alpha$.  
The short-range part of the correlation function, $C_{T_{\alpha b} T_{\alpha' b}}^{(1)}(t)$, can be obtained from the autocorrelation function $C_{\alpha \alpha'}(\textbf{r},t) = \langle E_{\alpha}(\textbf{r}, t') E_{\alpha'}^*(\textbf{r}, t' + t) \rangle$, where $E_{\alpha}(\textbf{r}, t)$ is the complex electric field inside the disordered medium illuminated by a wave in the incoming mode $\alpha$. If the time dependence of $E_{\alpha}(\textbf{r}, t)$ is due to the motion of scattering centers in the medium, $C_{\alpha \alpha'}(\textbf{r}, t)$ obeys the same diffusion equation as the average intensity $\overline{I}_{\alpha}(\textbf{r})$ but with $\chi^2 = \Delta {\textbf{q}}_{\alpha}^2 + \gamma^2(t) - 1/L_a^2$, where $\Delta \textbf{q}_{\alpha} = \textbf{q}_{\alpha} - \textbf{q}_{\alpha'}$ is the difference of transverse components $\textbf{q}$ of wavevectors $\textbf{k} = (\textbf{q}, k_z)$ of the incoming modes $\alpha$ and $\alpha'$.  $\gamma(t)$ describes decorrelation due to the motion of scattering centers and depends on the type and intensity of motion. For Brownian motion with a diffusion coefficient $D_B$, $\gamma^2(t) = 3 t/2t_0 \ell^2$ with $t_0 = 1/4 k^2 D_B$. Using the circular Gaussian statistics of $E_{\alpha}(\textbf{r}, t)$, we obtain
\begin{eqnarray}
&&C_{T_{\alpha b} T_{\alpha' b}}^{(1)}(t) =
\frac{1}{\overline{T}_{\alpha b}^2} \left(\frac{4 \pi D}{k^2 A}\right)^2
\left| \frac{\partial}{\partial z} C_{\alpha \alpha'}(\textbf{r}, t)\left|_{z = L} \right. \right|^2
\nonumber \\
&&\hspace{5mm} =
\delta_{\alpha \alpha'} \left| \frac{\sinh(\ell \sqrt{\gamma^2(t) - 1/L_a^2})}{\sinh(L \sqrt{\gamma^2(t) - 1/L_a^2})} \frac{\sin(L/L_a)}{\sin(\ell/L_a)} \right|^2
\label{appctab}
\end{eqnarray}
This correlation function is short-range in both time and space. It decays exponentially with time $t$ and it vanishes for $\alpha \ne \alpha'$. In the absence of amplification ($L_a \rightarrow \infty$), the correlation time of the fluctuations of $T_{\alpha b}$ is $t_c = \frac23 t_0 (\ell/L)^2$.

In order to calculate the long-range part of the correlation function, $C_{T_{\alpha b} T_{\alpha' b}}^{(2)}(t)$, we apply the well-known Langevin approach which we generalize to include both amplification (as in Ref.\ \cite{zyuzin95,burkov97}) and motion of scattering centers (as in Ref.\ \cite{skip01}). The fluctuation of intensity $\delta I_{\alpha}(\textbf{r}, t) = I_{\alpha}(\textbf{r},t) - \overline{I}_{\alpha}(\textbf{r})$ obeys a diffusion equation
\begin{eqnarray}
D \left( \nabla^2 + 1/L_a^2 \right) \delta I_{\alpha}(\textbf{r}, t) &=& \mathrm{div} \textbf{j}_{\alpha}(\textbf{r}, t)
\label{applang}
\end{eqnarray} 
with uncorrelated external Langevin currents 
\begin{eqnarray}
\overline{j_{\alpha}^{(i)}(\textbf{r}, t')
j_{\alpha'}^{(j)}(\textbf{r}, t'+t)} =
\delta_{ij} \frac{2 \pi \ell c^2}{3 k^2}
\left| C_{\alpha \alpha'}(\textbf{r}, t) \right|^2
\delta(\textbf{r} - \textbf{r}')
\nonumber \\
\label{appjext}
\end{eqnarray} 
Using the relation
\begin{eqnarray}
\delta T_{\alpha b}(t) &=& -\frac{4 \pi D}{k^2 A} \int d^2 \boldsymbol\rho \frac{\partial}{\partial z} \delta{I}_{\alpha}(\textbf{r},t)\left|_{z = L}\right.
\label{dtdi}
\end{eqnarray}
we obtain
\begin{eqnarray}
&&\overline{\delta T_{\alpha b}(t') \delta T_{\alpha' b}(t'+t)} = 
\frac{2 \pi \ell c^2}{3 k^2} \left( \frac{4 \pi D}{k^2 A} \right)^2 A 
\nonumber \\
&&\hspace{10mm} \times \frac{\partial}{\partial z} \frac{\partial}{\partial z'}
\int_0^L dz''
\frac{\partial}{\partial z''} {\tilde G}(z, z'')
\frac{\partial}{\partial z''} {\tilde G}(z', z'')
\nonumber \\
&&\hspace{10mm} \times \left| C_{\alpha \alpha'}(\textbf{r}'', t) \right|^2
 \left|_{z = z' = L}\right. 
\label{dtdt}
\end{eqnarray}
where
$\delta T_{\alpha b}(t) = T_{\alpha b}(t) - \overline{T}_{\alpha b}$
and ${\tilde G}(z, z'') = \int d^2 \boldsymbol\rho''
G(\textbf{r}, \textbf{r}'')$ is the $\textbf{q} = 0$ Fourier transform of $G(\textbf{r}, \textbf{r}'')$ with respect to $\boldsymbol\rho - \boldsymbol\rho''$.
Evaluating the integral in Eq.\ (\ref{dtdt}), dividing it by $\overline{T}_{\alpha b}^2$, and taking the limit of $L/\ell \gg 1$, we arrive at the following result:
\begin{eqnarray}
C_{T_{\alpha b} T_{\alpha' b}}^{(2a)}(t) &=&
\frac{1}{g} 
F_2(L/L_a, L \sqrt{\gamma^2(t) + \Delta \textbf{q}_{\alpha}^2})
\label{ct2a}
\end{eqnarray}
where $g = \frac43 N \ell/L$ is the dimensionless condutance and
\begin{eqnarray}
F_2(x, y) &=&
\frac{1}{4 x y^2 \sinh^2 \sqrt{y^2 - x^2}}
\nonumber \\
&\times& \left\{
\frac{x(2 y^2 - x^2)}{\sqrt{y^2 - x^2}}
\sinh 2 \sqrt{y^2 - x^2}
\right.
\nonumber \\
&-& \left.2 x y^2
- (y^2 - x^2) \sin 2x
\vphantom{\frac{x(2 y^2 - x^2)}{\sqrt{y^2 - x^2}}}\right\}
\label{f2}
\end{eqnarray}
For $y = 0$, corresponding to $t = 0$, Eq.\ (\ref{f2}) becomes
\begin{eqnarray}
F_2(x, y) &=&
\frac{1}{4} \left[
2 - \frac{\cotan x}{x} + \frac{1}{\sin^2 x}
\right]
\label{f20}
\end{eqnarray}

It is quite generally known that another contribution to the long-range correlation function $C_{T_{\alpha b} T_{\alpha' b'}}^{(2)}(t)$ exists \cite{vanrossum99,akkermans07}. By symmetry, this contribution depends on $\Delta \textbf{q}_b = \textbf{q}_b - \textbf{q}_{b'}$ in the same way as Eq.\ (\ref{ct2a}) depends on $\Delta \textbf{q}_a$. In our case, $b = b'$ and $\Delta \textbf{q}_b = 0$. Therefore, the final expression for the long-range correlation is
\begin{eqnarray}
C_{T_{\alpha b} T_{\alpha' b}}^{(2)}(t) &=&
\frac{1}{g} \left[ \vphantom{\sqrt{\gamma^2(t) + \Delta \textbf{q}_{\alpha}^2}}
F_2(L/L_a, \gamma(t)L) \right.
\nonumber \\
&+& \left. F_2(L/L_a, L \sqrt{\gamma^2(t) + \Delta \textbf{q}_{\alpha}^2})
\right]
\label{ct2}
\end{eqnarray}

The correlation function (\ref{ct2}) is long-range in space (it is of the same order for $\alpha = \alpha'$ and $\alpha \ne \alpha'$) and time (it decays only as $1/\sqrt{t}$ in the long-time limit). Because of the prefactor $1/g \ll 1$, it is, however, small in magnitude, at least as long as the amplification is weak ($x = L/L_a \ll 1$).

\section{Correlation function of $T_{\alpha b}(t)$ and $V_{b}(t)$}
\label{appctabvb}

The correlation function of $T_{\alpha b}(t)$ and $V_{b}(t)$ defined by Eq.\ (\ref{eq:CTabVb}) requires calculation of an average $\overline{\delta T_{\alpha b}(t') \delta V_b(t'+t)}$. Because of Eq.\ (\ref{eq:ConservationLaw}) and definitions of $T_b(t)$ and $R_b(t)$, this average can be represented as
\begin{eqnarray}
&&\overline{\delta T_{\alpha b}(t') \delta V_b(t'+t)}
\nonumber \\
&&= \sum_{\alpha'} \overline{\delta T_{\alpha b}(t') \delta T_{\alpha' b}(t'+t)}
 + \sum_{\beta} \overline{\delta T_{\alpha b}(t') \delta R_{\beta b}(t'+t)}
\nonumber \\
&&=
\overline{T}_{\alpha b} \sum_{\alpha'} \overline{T}_{\alpha' b} C_{T_{\alpha b} T_{\alpha' b}}(t)
+\overline{T}_{\alpha b} \sum_{\beta} \overline{R}_{\beta b} C_{T_{\alpha b} R_{\beta b}}(t)
\nonumber \\
\label{corr1}
\end{eqnarray}
Let us first consider the first sum. In the absence of amplification (i.e. for $L_a \rightarrow \infty$), $C_{T_{\alpha b} T_{\alpha' b}}(t)$ is known to have three distinct contributions denoted as $C^{(1)}$, $C^{(2)}$ and $C^{(3)}$ \cite{vanrossum99,akkermans07,berk94}. The first two of these were explicitly calculated in Appendix \ref{appctabtab}. $C^{(1)}$ is of order 1, but it vanishes for $\alpha' \ne \alpha$. $C^{(2)}$ is a factor $1/g$ smaller, but it is of similar magnitude for all $\alpha'$. Finally, $C^{(3)}$ is yet another factor $1/g$ smaller and, similarly to $C^{(2)}$, does not vanish for $\alpha' \ne \alpha$. We assume that this hierarchy of correlation functions does not change significantly in the presence of weak amplification (i.e. for $L_a$ finite but still $L_a \gg L$), which is indeed the case for $C^{(1)}$ and $C^{(2)}$, as can be seen from the results of Appendix \ref{appctabtab}. Then, when the different types of correlation functions are substituted into the first sum of Eq.\ (\ref{corr1}) and the summation is performed, they will yield $\overline{T}_{\alpha b}^2$, $(N/g) \overline{T}_{\alpha b}^2$ and $(N/g^2) \overline{T}_{\alpha b}^2$, respectively. Because in this paper we consider a situation when $N/g \sim L/\ell \gg 1$ and $g \gg 1$, the contribution of $C^{(2)}$ will dominate the result and the two other contributions can be neglected.

Suppose now that amplification is strong and we are close to the laser threshold (i.e. that $\Delta = 1 - L/\pi L_a \ll 1$). The behavior of equal-time correlation functions $C^{(i)}$ in this situation has been studied by Burkov and Zyuzin \cite{burkov97}: 
$C^{(1)} \sim \delta_{\alpha \alpha'}$,
$C^{(2)} \sim 1/g \Delta^2$ and
$C^{(3)} \sim 1/g^2 \Delta^4$.
Now $C^{(2)}$ will dominate the first sum of Eq.\ (\ref{corr1}) only if $\Delta \gg 1/\sqrt{g}$, otherwise the largest contribution will come from $C^{(3)}$. In the present paper we assume that the condition $\Delta \gg 1/\sqrt{g}$ is fulfilled, which also ensures that $C^{(2)}$ is always much smaller than $C^{(1)}$. We can therefore neglect all contributions to the correlation function $C_{T_{\alpha b} T_{\alpha' b}}(t)$ in Eq.\ (\ref{corr1}) other than $C^{(2)}$ for both weak and strong amplification, excluding only a narrow region in the vicinity of the laser threshold which requires a separate treatment and will not be considered here. Among the two terms contributing to $C^{(2)}_{T_{\alpha b} T_ {\alpha' b}}$ [see Eq.\ (\ref{ct2})], the first one $F_2(L/L_a, \gamma(t) L)$ will give the main contribution to the sum over $\alpha'$ in Eq.\ (\ref{corr1}) because it is independent of $\Delta \textbf{q}_a$, while the second term decays with $\Delta q_a$. We will neglect the second term of Eq.\ (\ref{ct2}) in the summation of Eq.\ (\ref{corr1}). 

Obviously, a similar reasoning applies to the second sum of Eq.\ (\ref{corr1}), except that $C^{(1)}$ correlation is always zero. $C_{T_{\alpha b} R_{\beta b}}(t)$ can be obtained along the same lines as $C_{T_{\alpha b} T_{\alpha' b}}(t)$ in Appendix \ref{appctabtab}.
Inserting the first term of Eq.\ (\ref{ct2}) and a similar result for $C_{T_{\alpha b} R_{\beta b}}(t)$ into Eq.\ (\ref{corr1}), we obtain
\begin{eqnarray}
&&C_{T_{\alpha b}V_{b}}(t) = \frac{1}{g} \times
\left\{
-\frac{x}{2 y^2} \cotan\frac{x}{2}
\right. 
\nonumber \\
&&\hspace{5mm} + \left. \frac{\coth\sqrt{y^2 - x^2}}{\sqrt{y^2 - x^2}}
\left( 1 - \frac{x^2}{2 y^2} \right) \right.
\nonumber \\
&&\hspace{5mm} - \left. \frac{1}{2 \sinh^2 \sqrt{y^2 - x^2}}
\left[ 1 - \frac{\sin x}{x y^2}(y^2 - x^2) \right]
\right\}
\hspace{10mm}
\label{resctabvb}
\end{eqnarray}
where $x = L/L_a$ and $y = \gamma(t) L$.
In particular, for $t = 0$ Eq.\ (\ref{resctabvb}) becomes
\begin{eqnarray}
C_{T_{\alpha b}V_{b}}(0) &=& \frac{1}{g} \times
\frac{1}{2 x \sin x} \left[ \frac{x}{\sin x}
\left( \frac{3}{2} + \cos x \right)
\right.
\nonumber \\
&-& \left. \frac{3}{2} \cos x - 1 \right]
\label{resctabvb0}
\end{eqnarray}

\section{Autocorrelation function of $V_{b}(t)$}
\label{appcvbvb}

Similarly to Appendix \ref{appctabvb}, we represent 
$\overline{\delta V_{b}(t') \delta V_b(t'+t)}$ as
\begin{eqnarray}
\overline{\delta V_{b}(t') \delta V_b(t'+t)} &=&
\sum_{\alpha, \alpha'} \overline{\delta T_{\alpha b}(t') \delta T_{\alpha' b}(t'+t)}
\nonumber \\
&+& \sum_{\beta, \beta'} \overline{\delta R_{\beta b}(t') \delta R_{\beta' b}(t'+t)}
\nonumber \\
&+& 2 \sum_{\alpha, \beta} \overline{\delta T_{\alpha b}(t') \delta R_{\beta b}(t'+t)}
\hspace{10mm}
\label{corr2}
\end{eqnarray}
For the same reasons and under the same conditions as in Appendix
\ref{appctabvb}, the main contributions to the three sums of Eq.\ (\ref{corr2}) come from $C^{(2)}$ correlation functions. As compared to Eq.\ (\ref{corr1}), Eq.\ (\ref{corr2}) contains a new correlation function $\overline{\delta R_{\beta b}(t') \delta R_{\beta' b}(t'+t)}$ which, however, can be calculated in the same way as the two others.
When divided by $\overline{V}_b^2$, Eq.\ (\ref{corr2}) yields
\begin{eqnarray}
&&C_{V_{b}V_{b}}(t) = \frac{1}{g} \times
\frac{1}{4 x y^2 \sqrt{y^2-x^2} \sin^2 \left(\frac{x}{2}\right)}
\nonumber \\
&&\hspace{5mm} \times \left\{
2 x \left[y^2 + (x^2 - y^2) \cos x \right]
\cotanh \sqrt{y^2-x^2} \right.
\nonumber \\
&&\hspace{5mm} - \left.
\frac{\sqrt{y^2-x^2}}{\sinh^2\sqrt{y^2-x^2}}
\left[ \vphantom{\sqrt{y^2-x^2}} 2 x y^2 \right. \right.
\nonumber \\
&&\hspace{5mm} + \left. \left. \left( x^2 - 2 y^2 + x^2 \cosh(2\sqrt{y^2-x^2})
\right) \sin x \right]
\right\}
\hspace{10mm}
\label{rescvbvb}
\end{eqnarray}
For $t = 0$ we obtain
\begin{eqnarray}
C_{V_{b}V_{b}}(0) &=& \frac{1}{g} \times
\frac{1}{16 x \left(\sin x \sin \frac{x}{2} \right)^2}
\left[
4 x (2 + \cos x)
\right.
\nonumber \\
&-& \left. 7 \sin x - 4 \sin 2x + \sin 3x
\right]
\label{rescvbvb0}
\end{eqnarray}


\end{document}